\documentclass[3p,11pt,sort&compress,review,number]{elsarticle}
\usepackage{graphicx}
\usepackage{amssymb}

\journal{Applied and Computational Harmonic Analysis}

\def\be{\begin{equation}}
\def\ee{\end{equation}}

\def\N{{\relax\ifmmode I\!\!N\else$I\!\!N$\fi}}
\def\R{{\relax\ifmmode I\!\!R\else$I\!\!R$\fi}}
\def\Z{{\relax\ifmmode Z\!\!\!Z\else$Z\!\!\!Z$\fi}}

\newtheorem{example}{Example}
\newtheorem{theorem}{Theorem}

\newtheorem{corollary}{Corollary}
\def\mod#1{~({\rm mod}~#1)} 

\begin{document}

\begin{frontmatter}


\title{Weyl-Heisenberg Spaces  for Robust Orthogonal Frequency Division Multiplexing}
\author[zoran]{Zoran Cvetkovi\'c\corref{cor1}}
\author[vincent]{Vincent Sinn}

\cortext[cor1]{Corresponding author.}

\address[zoran]{Department of Informatics, King's College London, Strand, London WC2R 2LS, UK,
E-mail: zoran.cvetkovic@kcl.ac.uk, Phone: + 44 20 7848 2858, Fax: + 44 20 7848 2932}
\address[vincent]{Telecommunications Laboratory, University of Sydney, NSW 2006, Australia, E-mail: cvsinn@ee.usyd.edu.au}




\begin{abstract}
Design of Weyl-Heisenberg sets of waveforms for robust orthogonal frequency division multiplexing (OFDM) 
has been the subject of a considerable volume of work.
In this paper,  a complete parameterization of orthogonal Weyl-Heisenberg sets and their corresponding 
biorthogonal sets is given. Several examples of Weyl-Heisenberg sets 
designed using this parameterization are presented, which in simulations
show  a high potential for enabling OFDM robust to frequency offset, timing mismatch, and
narrow-band interference.
\end{abstract}


\end{frontmatter}

\vspace*{\fill}

\pagebreak




\section{Introduction}

Orthogonal frequency division multiplexing (OFDM) is a communication
technique which is very effective in reducing or completely eliminating
intersymbol interference that arises in multipath propagation channels
\cite{chang,len,keller,nee,cherubini} and is rapidly
emerging as a technology of choice for wireless applications.
This robustness to multipath propagation is achieved by means of
frequency multiplexing which allows 
extending intersymbol interval beyond the duration of  
the channel impulse
response at the expense of a small level of redundancy.
To that end, a symbol sequence $a[n]$ is divided into subsequences, $a_k[i]=a[k+iN],~k=0,1,\ldots,N-1$, which are multiplexed in frequency and transmitted in {\sl frames} of $N$ symbols with {\sl frame intervals} of   $K$ samples, where $K>N$. The
transmitted signal thus has the form
\begin{equation}
s[n]=\sum_{i=-\infty}^{\infty} \sum_{k=0}^{N-1} a_{k}[i]
\varphi_k[n-iK]~,
\label{eq:dtmodel1}
\end{equation}
where waveforms  $\varphi_k[n],~k=0,1,\ldots,N-1,$ are commonly  complex
exponentials 
\begin{equation}
\varphi_k[n]= {1  \over \sqrt{N}} e^{j {2 \pi \over N}kn},~0 \leq n
\leq N-1~.
\label{eq:cpm}
\end{equation}
Subsequences
$a_k[i]$ are demultiplexed at the receiver by projecting $s[n]$
onto waveforms
\begin{equation}
\psi_k[n]={1\over \sqrt{N}}  e^{j {2 \pi \over N} kn},~0 \leq n \leq K-1~,
\label{eq:cpd}
\end{equation}
as 
$
a_k[i]=\langle \psi_k[n-iK],s[n]\rangle.
$
Each  waveform $\psi_k[n-iK]$ is orthogonal not only to all $\varphi_l[n-jK],~
k\neq l,~i\neq j$, but also to  all their delayed versions 
$\varphi_l[n-jK-d],$ for $d \leq K-N$.
In this manner
intersymbol interference due to multipath propagation is completely
eliminated if the impulse response of the channel does not exceed the {\sl guard interval}, $T_g=K-N$.
Alternatively, one can use waveforms $\psi_k[n]$ for the multiplexing and
waveforms $\varphi_k[n+N-K]$ for the demultiplexing. This alternative
method is known in communications as the
{\sl cyclic prefix} scheme \cite{peled}, 
since $s[n+iK]=s[n+N+iK],$ for $0 \leq n < K-N$,
whereas the design described by 
(\ref{eq:cpm}) and (\ref{eq:cpd})
is referred to as the {\sl zero-padding} scheme.
These two methods are equivalent for all considerations in this paper. 
Being so heavily optimized to
make the transmission robust to multipath propagation, the scheme 
is very sensitive to strong narrowband interference, frequency
offset and timing mismatch \cite{poll}. 
The sensitivity to frequency offset and
narrowband interference is caused by poor frequency 
localization 
of waveforms $\varphi_k[n]$ and $\psi_k[n]$, which is a consequence of
their short duration and sharp transitions. These sharp 
transitions are also responsible for the sensitivity to timing mismatch.

Optimal design of OFDM waveforms has been a very active area of research \cite{cherubini,ppv,tzannes,holte96,holte97,hass,hleiss,kozek,scaglione1,scaglione2,
cvetkovic99,bolcskei99,cvetkovic00,lin,cherubini02,siohan,benvenuto,strohmer,bolcskei03,phoong,
siclet,matz,jung}. The topic is currently  very 
relevant considering the role  OFDM is likely to play in 
the next generation of 
 wireless communication systems. One strand of work on OFDM waveform design 
is concerned with continuous-time waveforms, leading to many very 
insightful results, most importantly demonstrating that in the presence of
Doppler spread or frequency offset, waveforms which extend over several frame
intervals and attain high spectral containment achieve lower intersymbol (ISI)
and interchannel (ICI) interference than short rectangular waveforms, 
pointing out that a performance
can be improved if non-rectangular time-frequency transmission lattices are
used \cite{strohmer}, and concluding recently that with regard to practical
design excellent time-frequency localization of waveforms is the most
important requirement for low ISI/ICI \cite{matz}. The continuous-time
approach, on the other hand, has not provided yet closed form solutions, and 
optimal or nearly optimal waveforms are designed using numerical 
procedures which are quite challenging optimization problems per 
se \cite{jung}. 
Furthermore, for discrete-time implementation these 
waveforms are sampled and truncated, and that causes a non-negligible 
departure from the orthogonality and a degradation of frequency localization 
\cite{siohan}.

These insights and limitations of the continuous-time analysis 
motivate a purely discrete-time approach pursued in
\cite{hleiss,scaglione1,scaglione2,cvetkovic99,bolcskei99,cvetkovic00,siohan,bolcskei03,siclet}. Siohan, Siclet and
Lacaille  \cite{siohan}, as well as 
B\"{o}lcskei, Duhamel and Hleiss \cite{bolcskei03},
consider {\sl offset} OFDM with no redundancy, that is, the particular case 
when $K=N$, where owing to the offset multiplexing good time frequency 
localization of modulating waveforms is attainable; this is in contrast
to the OFDM with no offset  where good frequency localization
is impossible to achieve unless some redundancy is introduced. Since the 
primary source of robustness
of OFDM to multipath propagation is the  redundancy inserted by making the 
frame interval larger than the number of symbols in a frame \cite{scaglione1}, herewith we
focus on redundant OFDM schemes, {\sl i.e.}  cases with $K>N$.
In the direction of designing waveforms for redundant  OFDM robust to frequency offset and timing mismatch,
a straightforward approach  would be to perform optimization of their
time-frequency localization under required orthogonality constraints. This would,
however, be numerically very intensive and might have convergence problems in case of
long waveforms \cite{bolcskei99}. B\"{o}lcskei therefore proposes to orthogonalize well localized waveforms
using the discrete Zak transform   \cite{bolcskei99}; the method does not guarantee that the resulting 
orthogonal waveforms would still have good time-frequency localization, but it gave very good results in
examples with small number of channels and relatively high redundancies.  
More recently Siclet, Siohan and Pinchon proposed a method for finding orthogonal 
modulating waveforms \cite{siclet},  and used those solutions
to design waveforms via unconstrained optimization. The authors presented impressive 
design examples, but point out that there 
is no guarantee that their method works for arbitrary   $K$ and $N$ and 
suggest that low-redundancy systems  ($K$  close to $N$) may require 
a separate treatment. 

In this paper,  we present a complete parameterization,
{\sl i.e.} the complete set of solutions in a closed form, 
for  waveforms  
\begin{equation}
\varphi_k[n]= v[n] e^{j {2 \pi \over N}kn}~
\label{eq:varphi}
\end{equation}
which satisfy orthogonality conditions
\begin{equation}
\langle \varphi_k[n-iK],\varphi_l[n-jK]\rangle=\delta[k-l] \delta[i-j]~.
\label{eq:orth_fb1}
\end{equation}
The parameterization is valid for any arbitrary pair of parameters 
 $K$ and $N$,  $K \geq N$, and imposes no restrictions on the length of $v[n]$.
When there is 
redundancy in the system, $K>N$, given a modulating waveform $v[n]$ for which the
corresponding waveforms $\varphi_k[n]$ satisfy the orthogonality conditions in
(\ref{eq:orth_fb1}), there exist infinitely many solutions for a modulating waveform
$w[n]$ such that waveforms $\psi_k[n]$, 
\begin{equation}
\psi_k[n]= w[n] e^{j {2 \pi \over N}kn}~,
\label{eq:psi}
\end{equation}
are biorthogonal to corresponding waveforms $\varphi_k[n-iK]$,
\begin{equation}
\langle \psi_k[n-iK], \varphi_l[n-jK]\rangle=\delta[k-l]\delta[i-j]~.
\label{eq:biorth}
\end{equation}
These waveforms $\psi_k[n]$ can also be used for perfect demultiplexing
instead of waveforms $\varphi_k[n]$ themselves. 
We give  a complete parameterization of such modulating waveforms $w[n]$,
as this additional degree of design 
freedom might result in possible further improvements in system performance; 
the zero-padding and cyclic prefix schemes are  particular cases of the 
complete set of solutions explored here, obtained when
$v[n]$ is the $N$-sample rectangular waveform and $w[n]$ 
is the $K$-sample rectangular waveform. 
The parameterization
of orthogonal waveforms $\varphi_k[n]$ presented here is a generalization of
the idea proposed by Hleiss, Duhamel and Charbit \cite{hleiss}, inspired by work
of the first author and Vetterli on Weyl-Heisenberg frames in $\ell^2(\Z)$ 
\cite{cvetkovic98}. Hleiss, Duhamel and Charbit observe that the parameterization
of tight Weyl-Heisenberg frames in \cite{cvetkovic98} can be  used to find
solutions for OFDM  waveforms and present some design example, but point out 
that
proving a general result for arbitrary $K$ and $N$ is quite intricate \cite{hleiss}. 
The  parameterization of OFDM waveforms and corresponding biorthogonal
demultiplexing waveforms given here were previously outlined by the first
author in conference publications \cite{cvetkovic99,cvetkovic00}. 
Considering the increasingly important role OFDM is 
playing  in  wireless communication technologies, there is a need for 
precise, complete,  and detailed  presentation of those solutions and provided it in this paper.

One may argue that imposing the orthogonality conditions in 
(\ref{eq:orth_fb1}) may preclude  exploring the complete set of 
modulating waveforms, since this orthogonality is not 
necessary for perfect demultiplexing.
As long as waveforms 
$\varphi_k[n-iK]$
are linearly independent, which is  satisfied under very mild
conditions on $v[n]$, there  exists a modulating waveform $w[n]$ such that its
modulated translates 
$\psi_k[n-iK]$ are biorthogonal to waveforms $\varphi_k[n-iK]$ according 
to (\ref{eq:biorth}).
Such waveforms $\psi_k[n]$ can  be used for perfect demultiplexing in
channels with no multipath propagation even when waveforms $\varphi_k[n-iK]$ 
are not mutually orthogonal, and 
this additional design freedom can be used
to achieve a better joint time-frequency localization of $v[n]$ and $w[n]$ and thus reduce ISI/ICI. 
Besides, both orthogonality and biorthogonality are lost in multipath
channels except in the particular case of zero-padding and cyclic 
prefix schemes. 
The idea of biorthogonal frequency division multiplexing (BFDM) in the continuous-time case was explored in 
detail in \cite{kozek} and \cite{matz}.  The biorthogonal waveforms designed to minimize
ISI/ICI in \cite{kozek} are in fact
very close to orthogonal, and the conclusion of the theoretical analysis
in \cite{matz} was that waveforms which minimize the interference should
be very close to orthogonal. A question which arises is whether orthogonal
rather than biorthogonal multiplexing is in fact optimal in terms of minimizing
the interference and that design examples presented in \cite{kozek} and 
\cite{matz} were close to orthogonal not because
optimal waveforms are biorthogonal but because the numerical procedures 
approached optimal solutions but did not reach them exactly. Orthogonal multiplexing is also
optimal in terms of robustness to additive white Gaussian noise, as it has been  established 
already by Shannon \cite{shannon}, and also proved more recently by 
Kozek and Molisch \cite{kozek}.  This motivates 
the focus of this work on orthogonal waveforms, and additional design
freedom is provided by solutions for biorthogonal demultiplexing, should there
be some merit to it as it happens in the case of the cyclic prefix scheme, or
should the transmultiplexer be optimized according to different noise and channel
statistics as explored by Scaglione, Giannakis and Barbarossa in the case
of waveforms up to $K$ samples long \cite{scaglione1}.


The parameterization of Weyl-Heisenberg sets for OFDM
given here is based on a polyphase representation of
OFDM transmultiplexer, which is reviewed first in Section II . Further parameterization details, the particular  
form which OFDM waveforms have and the relationship between orthonormal Weyl-Hesinberg sets and
tight Weyl-Heisenberg frames are discussed in Section III. Some waveform design 
issues are discussed in Section IV. Design examples and simulation results
are presented in Section V.

\section{Polyphase Representation of OFDM}

\subsection{Transmultiplexer Polyphase Representation}

\begin{figure}[thb]
\centerline{
\includegraphics[width=0.65\textwidth]{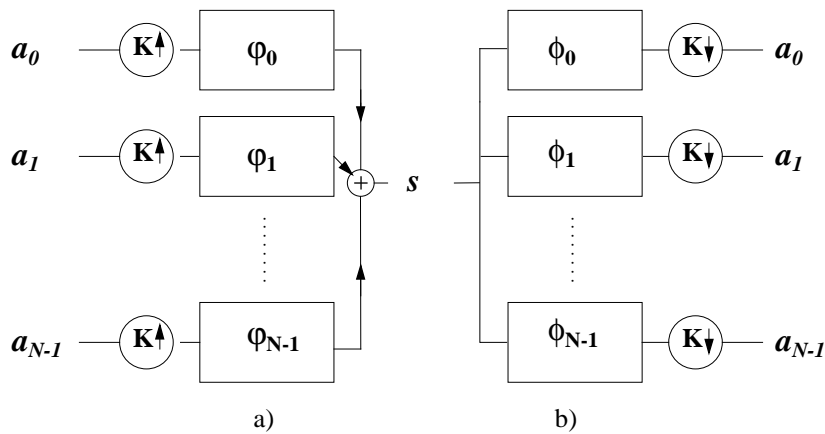}}
\caption{\sl Filter banks for implementation of $N$-channel OFDM with
$K$-sample frame interval. a) Multiplexer. b) Demultiplexer.}
\label{fig:mxdmx}
\end{figure}

An  OFDM multiplexer has the structure of a multirate synthesis filter bank.
The filter bank implementing an $N$-channel
multiplexer with $K$-sample  frame interval,
as shown in Figure \ref{fig:mxdmx}a),
performs $K$-point upsampling of
input sequences, followed by linear  filtering in each
of its channels.
 The multiplexer output is  given by
$\displaystyle{
s[n]=\sum_{i=-\infty}^{\infty}\sum_{k=0}^{N-1}a_k[i] \varphi_k[n-iK],
}$
where $a_k[i]$ and $\varphi_k[n]$ denote the input sequence
and the impulse response of the filter in the channel $k$, respectively.
The treatment of orthogonal frequency division multiplexing  in
this paper is based on the techniques of polyphase analysis of
multirate  systems \cite{ppv,martin}.

For the analysis of multiplexers with $K$-point
upsampling it is convenient to represent the multiplexed signal $s[n]$
in terms of its $K$ polyphase components as
$S(z)=\sum_{l=0}^{K-1}S_l(z^K)z^{-l},$  where $S_l(z)=\sum_{n=-\infty}^{\infty}s[l+nK]z^{-n}.$
The polyphase components of $s[n]$ are related to the input sequences as
$
[S_0(z) \ldots S_{K-1}(z)]^T = {\bf M}(z)[A_0(z)
\ldots A_{N-1}(z)]^T,
$
where ${\bf M}(z)$ is the $K \times N$ 
multiplexer polyphase matrix, and $A_k(z)$ is the $z$-transform
of the sequence $a_k[i]$.
Entries of ${\bf M}(z)$ are given by\footnote{Throughout the paper, matrix rows and columns will be
indexed starting with zero.}:
$[{\bf M}(z)]_{l,k}=\sum_{n=-\infty}^{+\infty}\varphi_k[l+nK]z^{-n},
~0\leq l \leq K-1,~0 \leq k \leq N-1.$

The demultiplexer 
has the structure of an $N$-channel analysis filter bank,  as shown in
Figure \ref{fig:mxdmx}b). 
The sequence $b_k[i]$ produced  in the channel $k$ of the demultiplexer
is given by
$
b_k[i]=\sum_{n=-\infty}^{\infty} \phi_k[iK-n]s[n],
$
where $\phi_k[n]$ is the impulse response of the filter in that channel.  For $\phi_k[n]=\psi_k^\ast[-n]$ the
output sequence becomes
$
b_k[i]=\langle \psi_k[n-iK],s[n]\rangle .
$
To simplify the notation, all following 
considerations will be expressed in terms of  impulse responses $\phi_k[n]$ rather than waveforms $\psi_k[n]$.
Output sequences of the demultiplexer are related to the
input signal $s[n]$ as
$
[B_0(z) \ldots B_{N-1}(z)]^T={\bf D}(z)
[S_0(z) \ldots S_{K-1}(z)]^T,
$
where ${\bf D}(z)$ is the polyphase representation of the
demultiplexer, given by
$[{\bf D}(z)]_{k,l}=\sum_{n=-\infty}^{+\infty}\phi_k[-l+nK]z^{-n},
~0\leq k \leq N-1,~0 \leq l \leq K-1,
$
and $B_k(z)$ is the $z$-transform of the sequence $b_k[i]$.
Sequences $b_k[i]$ 
are perfectly demultiplexed  sequences $a_k[i]$
if and only if the polyphase demultiplexer matrix is a left inverse of
the polyphase multiplexer matrix, ${\bf D}(z){\bf M}(z)={\bf I},$
where ${\bf I}$ denotes the $N \times N$ identity matrix,
or  equivalently if and only if 
the following biorthogonality relationships
hold \cite{ppv,martin}:
\begin{equation}
\sum_n \phi_k[jK-n]\varphi_l[n-iK]=\delta[k-l] \delta[i-j]~.
\label{eq:biorth_fb}
\end{equation}
When the multicarrier scheme is redundant, $K>N$, and
waveforms $\varphi_k[n-iK],~k=0,1,\ldots N-1,~i \in\Z$, are linearly 
independent, the polyphase multiplexer matrix  ${\bf M}(z)$ has infinitely many
left inverses ${\bf D}(z)$, that is, there exist infinitely
many filters $\phi_k[n],~k=0,1,\ldots,N-1,$ which can be used for perfect 
demultiplexing.
In the case when the multiplexer
waveforms $\varphi_k[n]$ satisfy the orthogonality conditions  in
(\ref{eq:orth_fb1}), one solution for filters which would achieve
perfect demultiplexing are filters which are
complex-conjugated time-reversed versions of the multiplexer waveforms,
$\phi_k[n]=\varphi^\ast_k[-n]$.
In that case, the polyphase representation of the demultiplexer
is the Hilbert adjoint of the multiplexer polyphase representation,
${\bf D}(z)={\bf \tilde M}(z),$\footnote{${\bf \tilde{M}}(z)$
denotes the matrix obtained by transposing  ${\bf M}(z)$, conjugating all the
coefficients of the polynomials in ${\bf M}(z),$ and replacing $z$
by $z^{-1}$.} which further means that the orthogonality
conditions in (\ref{eq:orth_fb1}) are equivalent to the paraunitariness of
${\bf M}(z)$,
${\bf \tilde M}(z){\bf M}(z)={\bf I}.$
Towards finding a parameterization of OFDM pulse shapes,  in the
next subsection we consider the particular form ${\bf D}(z)$ and ${\bf M}(z)$ 
have in the case of OFDM.

\subsection{Polyphase Representation of OFDM Transmultiplexer}\label{ss:ofdm_pp}

Waveforms in an OFDM multiplexer are modulated complex exponentials,
$\varphi_k[n]= v[n] e^{j {2 \pi \over N}kn},$ and then  
the polyphase multiplexer representation 
has a particular form established by the following theorem.
\begin{theorem}\label{lemma}
Consider an $N$-channel OFDM multiplexer with $K$-sample frame interval, 
based on a modulating waveform $v[n]$. Let $M$ be the least common multiple of $K$ and $N$, and let $V_j(z), ~j=0,1,\ldots,M-1,$ be the components of the $M$-component polyphase representation of $v[n]$,  
\begin{equation}\label{eq:mpolyphase}
V_j(z) =  \sum_{n = - \infty}^{\infty} v[j+nM]z^{-n}~. 
\end{equation}
The polyphase representation of this multiplexer has the form
\begin{equation}
{\bf M}(z)={\bf V}(z) {\bf F}_N~,
\label{eq:mvz}
\end{equation}
where 
${\bf F}_N$ is the $N$-point discrete-Fourier transform matrix,
$[ {\bf F}_N]_{m,n}=e^{j{2 \pi \over N}mn},~0 \leq m \leq N-1,
~0 \leq n \leq N-1,$ and
${\bf V}(z)$ is the $K \times N$ matrix of the polyphase components of $v[n]$ given by
\begin{equation}\label{eq:vform1}
[{\bf V}(z)]_{l,m}= \left\{ \begin{array}{cc}
z^{-p}V_{pK+iP+r}(z^J)~, & (l,m)=(iP+r,jP+r)\\
0~, & {\rm otherwise}
\end{array} \right. ,
\end{equation}
$i=0,1,\ldots,L-1$, $j=0,1,\ldots,J-1$, $r=0,1,\ldots,P-1$,  where $P$ is the greatest common divisor of $N$ and $K$,
$J=N/P$, $L=K/P$ and $p$ is the integer such that
\begin{equation}\label{eq:ijcongr}
j \equiv pL+i \mod{J}~.
\end{equation}
\end{theorem}
{\bf Proof:} To show that ${\bf M}(z)$ has this particular form,
consider
$$[{\bf M}(z)]_{l,k}=\sum_{n=-\infty}^{+\infty}\varphi_k[l+nK]z^{-n}=
\sum_{n=-\infty}^{+\infty}v[l+nK]e^{j {2 \pi \over N}k(l+nK)}z^{-n}~.$$
By representing  $n$ as $n=qJ+p$, $0 \leq p \leq J-1$, $q \in
{\Z}$, we obtain
$$[{\bf M}(z)]_{l,k}=\sum_{p=0}^{J-1} e^{j {2 \pi \over N}k(l+pK)}z^{-p}
\sum_{q=-\infty}^{\infty} v[l+pK+qJK]z^{-qJ}.
$$
This gives
$$[{\bf M}(z)]_{l,k}=\sum_{p=0}^{J-1} e^{j {2 \pi \over
N}k(l+pK)}z^{-p}V_{pK+l}(z^J).
$$
Consequently, we obtain
$$[{\bf M}(z)]_{l,k}=\sum_{p=0}^{J-1} e^{j {2 \pi \over
N}km(l,p)}z^{-p}V_{pK+l}(z^J)~,
$$
where $m(l,p) \equiv pK+l ~({\rm mod}~N)$. Hence, the $l$-th row of ${\bf V}(z)$ has $J$ nonzero entries,
$z^{-p}V_{pK+l}(z^J)$, $p=0,1,\ldots, J-1$,  positioned so that 
$z^{-p}V_{pK+l}(z^J)$ is in the column 
$m \equiv pK+l \mod{N}$, or equivalently, in the column $m$ for which there exists an integer $q$ such that
\begin{equation}
qN+m=pK+l~.
\label{eq:mlcongr}
\end{equation}
By representing $N=JP$ and $K=LP$ the equality in (\ref{eq:mlcongr}) becomes 
$qJP+m=pLP+l$ which is equivalent to $m \equiv  l \mod{P}$. 
Hence nonzero entries of ${\bf V}(z)$ appear only at locations $(l,m)=(iP+r,jP+r)$, and the entry at 
$(l,m)=(iP+r,jP+r)$ is $z^{-p}V_{pK+l}=z^{-p}V_{pK+iP+r}$, where $p$ is the integer which satisfies (\ref{eq:mlcongr}). By expressing
$l=iP+r$, $m=jP+r$, $N=JP$ and $K=LP$ the congruence in  (\ref{eq:mlcongr}) reduces to $qJ+j=pL+i$ which is
equivalent to $j \equiv pL+i \mod{J}$. 
\hspace*{\fill}$\square$

Thus only entries $[{\bf V}(z)]_{l,m}$ at locations $[{\bf V}(z)]_{iP+r,jP+r}$, $i=0,1,\ldots,L-1$, $j=0,1,\ldots, J-1$, $r=0,1,\ldots,P-1$ are different from zero.
Hence each row  of ${\bf V}(z)$  has $J$ nonzero components, and each column has $L$ nonzero entries.
When $N$ and $K$ are coprime $J=N$ and $L=K$, and consequently ${\bf V}(z)$ is a full matrix.
The structure of ${\bf V}(z)$ is illustrated in the following example. 

\begin{example} \label{ex:v64}
{\em In the case $N=4$, $K=6$, ${\bf V}(z)$ has the form
$$
{\bf V}(z)= \left[
\begin{array}{cccc}
              V_0(z^2) & 0 & z^{-1}V_6(z^2) & 0 \\
              0 & V_1(z^2) & 0  & z^{-1}V_7(z^2)\\
              z^{-1}V_8(z^2) & 0 & V_2(z^2) & 0\\
              0 & z^{-1}V_9(z^2) & 0 & V_3(z^2) \\
              V_4(z^2) & 0 & z^{-1}V_{10}(z^2) & 0 \\
              0 &  V_5(z^2) & 0 & z^{-1}V_{11}(z^2)
\end{array}
\right].
$$} 
\hfill$\square$
\end{example}

Filters of a corresponding demultiplexer are also windowed complex
exponentials, $\phi_k[n]= w[n] e^{j {2 \pi \over N}kn}.$
Analogously to the derivation of the polyphase
representation of the multiplexer, it can be shown that the polyphase 
matrix of
the demultiplexer has the form ${\bf D}(z)={\bf F}^\ast_N {\bf W}(z),$
where ${\bf F}^\ast_N$ is the complex-conjugated transpose of ${\bf F}_N$,
and ${\bf W}(z)$ is an $N \times K$ matrix of  polyphase
components of  $w[n]$.
To specify ${\bf W}(z)$, consider the
following set of polyphase components of $w[n]$:
$
W_j(z) =  \sum_{n = - \infty}^{\infty} w[-j+nM]z^{-n},~j=0,1,\ldots M-1.
$
In a manner analogous to the
derivation of ${\bf V}(z)$, it can be shown that 
\begin{equation}
[{\bf W}(z)]_{l,m}=\left\{ \begin{array}{cc} 
z^pW_{pK+jP+r}(z^J)~, & 
(l,m)=(iP+r,jP+r) \\
0~, & {\rm otherwise}
\end{array}
\right. ~,
\label{eq:wform}
\end{equation}
where $i \equiv pL+j \mod{J}$.

Expressed in terms of  ${\bf V}(z)$ and  ${\bf W}(z)$,
the condition for perfect transmultiplexing is
that ${\bf W}(z)$ is (modulo a multiplicative constant) a left inverse of ${\bf V}(z)$:
${\bf W}(z){\bf V}(z)=(1/ N) {\bf I}.$
The left inverse of ${\bf V}(z)$
is not unique when ${\bf V}(z)$  is a $K \times N$ matrix where $K>N$,
hence, given a multiplexing  waveform  $v[n]$, the corresponding
waveform $w[n]$ which achieves perfect demultiplexing is not unique.
Based on these polyphase description of the multiplexer and the demultiplexer, in the next subsection
we will provide a complete parameterization of orthogonal Weyl-Heisenberg sets in $\ell^2(\Z)$. 
This polyphase multiplexer description also  leads to fast algorithms for implementation of OFDM based on long modulating
waveforms  \cite{cvetkovic99}.

\section{Parameterization of Orthonormal Weyl-Heisenberg Sets in $\ell^2(\Z)$}
\label{sec:parameterization}

\subsection{Orthogonal Weyl-Heisenberg Sets}

From considerations in the previous section, it follows that modulating waveforms satisfy the
orthogonality conditions  in (\ref{eq:orth_fb1}) if and only if
the corresponding matrix ${\bf V}(z)$, as defined in (\ref{eq:mvz}), is paraunitary:
\begin{equation}
{\bf \tilde V}(z){\bf V}(z)={1 \over  N}{\bf I}~.
\label{eq:parav}
\end{equation}
From the particular structure of ${\bf V}(z)$, as described in the previous subsection it follows 
that  ${\bf V}(z)$ satisfies (\ref{eq:parav}) if and only if a particular set of its submatrices satisfy the
condition.
This is illustrated by the following example.

\begin{example}
{ In the case given in Example \ref{ex:v64},  ${\bf V}(z)$
is paraunitary if and only if its  submatrices
\begin{equation}
{\bf V}_0(z)=\left[
\begin{array}{cc}
              V_0(z^2) & z^{-1}V_6(z^2) \\
              z^{-1}V_8(z^2) & V_2(z^2)\\
              V_4(z^2) & z^{-1}V_{10}(z^2)
\end{array}
\right]
~~~
{\bf V}_1(z)=\left[
\begin{array}{cc}
              V_1(z^2)  & z^{-1}V_7(z^2)\\
              z^{-1}V_9(z^2) &  V_3(z^2) \\
              V_5(z^2) & z^{-1}V_{11}(z^2)
\end{array}
\right]
\end{equation}
are paraunitary.
Further, matrices ${\bf V}_i(z),~i=0,1$  are paraunitary
if and only if matrices 
\begin{equation}
{\bf V}^o_0(z)=\left[
\begin{array}{cc}
              V_0(z) & V_6(z) \\
              V_8(z) & zV_2(z)\\
              V_4(z) & V_{10}(z)
\end{array}
\right]
~~~
{\bf V}^o_1(z)=\left[
\begin{array}{cc}
              V_1(z)  & V_7(z)\\
              V_9(z) &  zV_3(z) \\
              V_5(z) &   V_{11}(z)
\end{array}
\right]
\end{equation}
are paraunitary, where ${\bf V}^o_i(z^2)={\rm diag}(1,z,1){\bf V}_i(z)
{\rm diag}(1,z),~i=0,1$.
\hfill$\square$
}
\end{example}

 A sufficient and necessary condition for the orthogonality relationships in (\ref{eq:orth_fb1}) is established by the following theorem, from which a complete parameterization of orthonormal Weyl-Heisenberg sets in $\ell^2(\Z)$ follows immediately.

\begin{theorem}\label{maintheorem}
Consider a Weyl-Heisenberg set $\Phi=\{\varphi_{k,i}:\varphi_{k,i}[n]=v[n-iK]e^{j {2 \pi \over N}k(n-iK)}\}_{k \in \Z_N, i \in \Z}$. 
Let $M$ be the least common multiple of $K$ and $N$, and let $V_j(z), ~j=0,1,\ldots,M-1,$ be the components of the $M$-component polyphase representation of $v[n]$ as given 
by (\ref{eq:mpolyphase}). $\Phi$ is an orthonormal set if and only if the following 
matrices are paranunitary:
\begin{equation}\label{eq:vrform}
[{\bf V}_r^o(z)]_{i,j}=
z^{n(i,j)}V_{p(i,j)K+iP+r}(z)~,
\end{equation}
$i=0,1,\ldots,L-1$, $j=0,1,\ldots,J-1$, $r=0,1,\ldots,P-1$,  where $P$ is the greatest common divisor of $N$ and $K$,
$J=N/P$, $L=K/P$, $p(i,j)$ is the integer such that
\begin{equation}\label{eq:ijcongr}
j \equiv p(i,j)L+i \mod{J}~.
\end{equation}
and $n(i,j) \in \{0,1\}$ is given by $n(i,j)=[p(i,0)+p(0,j)-p(i,j)]/J$.
\end{theorem}
{\bf Proof:}
It follows from (\ref{eq:vform1}) that ${\bf V}(z)$ in (\ref{eq:mvz}) is paraunitary if and only if its submatrices 
${\bf V}_r(z)$, $r=0,1,\ldots,P-1,$
\begin{equation}
[{\bf V}_r(z)]_{i,j}=[{\bf V}(z)]_{iP+r,jP+r}=z^{-p(i,j)}V_{p(i,j)K+iP+r}(z^J)~,
\end{equation}
where $p(i,j)$ is the integer which satisfies $j \equiv p(i,j)L+i \mod{J}$,
are paraunitary. Further,  ${\bf V}_r(z)$ is paraunitary if and only if
${\bf V}_r'(z)={\bf D}_R(z) {\bf V}_r(z){\bf D}_C(z)$ is paraunitary, where 
$${\bf D}_L(z)={\rm diag}\left(z^{p(0,0)}, z^{(p(0,1)},\ldots,z^{p(0,L-1)}\right)~,~
{\bf D}_C(z)={\rm diag}\left(z^{p(0,0)}, z^{(p(1,0)},\ldots,z^{p(J-1,0)}\right)~.$$
Matrices ${\bf V}_r'(z)$ have the form
$$[{\bf V}_r'(z)]_{i,j}=z^{p(i,0)+p(0,j)-p(i,j)}V_{p(i,j)K+iP+r}(z^J)~.$$
Since $j \equiv p(i,j)L+i \mod{J}$, there exists an integer $q(i,j)$ such that 
$$q(i,j)J+j=p(i,j)L+i~.$$
This implies that $(q(i,0)+q(0,j)-q(i,j))J=(p(i,0)+p(0,j)-p(i,j))L$, and since $J$ and $L$ are coprime,
$p(i,0)+p(0,j)-p(i,j)$ must be a multiple of $J$. Furthermore, since $0 \leq p(i,j) \leq J-1$, the sum
$p(i,0)+p(0,j)-p(i,j)$ must be either $0$ or $J$. Hence, 
${\bf V}_r'(z)={\bf V}_r^o(z^J)$, where ${\bf V}_r^o(z)$ is as given in the statement of the theorem and
${\bf V}(z)$ is paraunitary if and only if all ${\bf V}_r(z)$ are paraunitary.
\hfill$\square$

\begin{corollary}\label{maincorollary}
{ The $M={\rm LCM}(K,N)$ polyphase components of a waveform $v[n]$
for which the Weyl-Heisenberg set $\Phi=\{\varphi_{k,i}:\varphi_{k,i}[n]=v[n-iK]e^{j {2 \pi \over N}k(n-iK)}\}_{k \in \Z_N, i \in \Z}$ 
is orthonormal  
are  up to time delays entries of $P={\rm GCD}(N,K)$   paraunitary matrices
of size $L \times J$, and vice versa, entries of $P$ arbitrary 
paraunitary  $L \times J$ matrices are up to time delays the polyphase components of
a waveform   $v[n]$ for which $\Phi$ is an orthonormal set.}
\end{corollary}

 Parameterizations of paraunitary
matrices have been previously studied in the filter bank literature
\cite{ppv,martin}, and these combined with the special form of ${\bf V}_r^o(z)$
described by (\ref{eq:vrform}) provide a complete parameterization of 
OFDM modulating waveforms.  Details of this parameterization are illustrated by
the following example.

\begin{example}\label{ex:v128_160pp}
{Consider a Weyl-Heisenberg set  $\Phi$ underlying OFDM with $N=128$-channels and $K=160$-sample frame interval. The polyphase multiplexer matrix is paraunitary if and only if the following 
submatrices of the corresponding matrix ${\bf V}(z)$ are paraunitary:
$$
{\bf V}_i(z)= \left[
\begin{array}{cccc}
              V_{0\cdot 32+i}(z^4) & z^{-1}V_{5\cdot 32+i}(z^4) & 
z^{-2}V_{10 \cdot 32+i}(z^4) & z^{-3}V_{15 \cdot 32 +i}(z^4)  \\
              z^{-3}V_{16 \cdot 32 +i}(z^4) & V_{1 \cdot 32 +i}(z^4) &
              z^{-1}V_{6\cdot 32 +i}(z^4) & z^{-2}V_{11\cdot 32 +i}(z^4) \\
             z^{-2} V_{12 \cdot 32 +i}(z^4) & z^{-3}V_{17 \cdot 32 +i}(z^4) &
              V_{2\cdot 32 +i}(z^4) & z^{-1}V_{7 \cdot 32 +i}(z^4) \\
              z^{-1}V_{8 \cdot 32 +i}(z^4) & z^{-2}V_{13 \cdot 32 +i}(z^4) & 
z^{-3}V_{18 \cdot 32 +i}(z^4) & V_{3\cdot 32 +i}(z)  \\
              V_{4 \cdot 32 +i}(z^4) & z^{-1}V_{9 \cdot 32 +i}(z^4) & 
z^{-2}V_{14 \cdot 32 +i}(z^4) & z^{-3}V_{19 \cdot 32 +i}(z^4)
\end{array}
\right],~~~i=0,1,\ldots 31.
$$
To obtain matrices of polyphase components of $v[n]$ with only unit delays,
matrices ${\bf V}_i(z)$ are multiplied from the right by 
${\rm diag}(1,z,z^2,z^3)$, the inverse delays of the first row of 
${\bf V}_i(z)$, and from the 
left by ${\rm diag}(1,z^3,z^2,z,1)$, the inverse delays of the first column
of ${\bf V}_i(z)$. 
This procedure
gives matrices
$${\bf V}^o_i(z^4)={\rm diag}(1,z^3,z^2,z,1){\bf V}_i(z)
{\rm diag}(1,z,z^2,z^3)~,$$ 
where
\begin{equation}
{\bf V}^o_i(z)= \left[
\begin{array}{cccc}
              V_{0\cdot 32+i}(z) & V_{5\cdot 32+i}(z) & V_{10 \cdot 32
              +i}(z) & V_{15 \cdot 32 +i}(z)  \\
              V_{16 \cdot 32 +i}(z) & zV_{1 \cdot 32 +i}(z) &
              zV_{6\cdot 32 +i}(z) & zV_{11\cdot 32 +i}(z) \\
              V_{12 \cdot 32 +i}(z) & V_{17 \cdot 32 +i}(z) &
              zV_{2\cdot 32 +i}(z) & zV_{7 \cdot 32 +i}(z) \\
              V_{8 \cdot 32 +i}(z) & V_{13 \cdot 32 +i}(z) & V_{18
              \cdot 32 +i}(z) & zV_{3\cdot 32 +i}(z)  \\
              V_{4 \cdot 32 +i}(z) & V_{9 \cdot 32 +i}(z) & V_{14
              \cdot 32 +i}(z) & V_{19 \cdot 32 +i}(z)
\end{array}
\right],~~~i=0,1,\ldots 31.
\label{eq:vio}
\end{equation}
It follows that ${\bf V}_i(z)$ is paraunitary if and only if ${\bf V}^o_i(z)$
is paraunitary, and consequently, ${\bf V}(z)$ is paraunitary if and only
if all ${\bf V}^o_i(z),~i=0,1,\ldots,31$ are paraunitary.

A modulating waveform for OFDM with $N=128$ and $K=160$ is then obtained as follows: i) start with an arbitrary set of 
${\rm GCD}(N,K)=32$ paraunitary matrices 
${\bf V}^o_i(z),~i=0,1,\ldots,31$ of size 
$L \times J=5\times 4$,
and express coefficients of polynomials in ${\bf V}^o_i(z)$ in terms of 
free parameters, {\sl e.g.} angles of Given's
rotations (see \cite{ppv,martin});  ii) interleave entries
of ${\bf V}^o_i(z)$ according to (\ref{eq:vio}) to form $v[n]$. 
An example of waveform obtained by applying an unconstrained optimization procedure to a solution $v[n]$ obtained 
in this manner is shown in Figure 
\ref{fig:w160_1024}a).
Matrices  ${\bf V}^o_i(z)$ corresponding to this waveform are 
paraunitary matrices of zero degree monomials
(i.e. unitary scalar matrices), so $v[n]$  has $640$ nonzero taps.
However, the delay elements in ${\bf V}^o_i(z)$, that appear with
$ V_{1 \cdot 32 +i}(z)$, $V_{2\cdot 32+i}(z)$, $V_{3\cdot 32+i}(z)$,
$V_{6\cdot 32+i}(z)$, $V_{7\cdot 32+i}(z)$, and $V_{11\cdot 32+i}(z)$,
create zero taps, so the total waveform
length is $L_v=1024$. The existence of these zero taps cannot be avoided in
OFDM design, except when  $K/N$ is integer.

Another orthogonal OFDM waveform is the $N$ samples long rectangular waveform.
The corresponding polyphase matrices ${\bf V}^o_i(z)$ are all equal, and  are given by
\begin{equation} \label{eq:cpv}
{\bf V}^o_i(z)= {1 \over \sqrt{128}}\left[
\begin{array}{c}
                  {\bf  0}  \\
                   {\bf I}
\end{array}
\right],~~~i=0,1,\ldots 31~,
\end{equation}
where ${\bf 0}=[0~0~0~0]$ and ${\bf I}$ is the $4 \times 4$ identity matrix. }
\hfill $\square$
\end{example}

\subsection{Biorthogonal Demultiplexing Waveforms}

To parameterize all solutions for the 
demultiplexing waveforms, or equivalently, the complete set of inverses of ${\bf V}(z)$,  consider a paraunitary ${\bf V}(z)$ and the
associated paraunitary matrices ${\bf V}^o_i(z),~i=0,1,\ldots,N/J$.
The size of each ${\bf V}^o_i(z)$ is $L\times J$, where $L>J$.
Each  matrix ${\bf V}^o_i(z)$ can be completed to a square $L\times
L$ paraunitary matrix
${\bf V}_i^s(z)=[{\bf V}^o_i(z)~ {\bf V}_i^c(z)]$ \cite{alan}.
Let ${\bf U}_i(z)$ be an $L\times J$ matrix given by
\begin{equation}
{\bf U}^o_i(z)={\bf V}_i^s(z) \left[ \begin{array}{c}
{\bf I} \\
{\bf A}_i(z) \end{array} \right] ~,
\end{equation}
where ${\bf I}$ is the $J \times J$ identity matrix and ${\bf A}_i(z)$ is an $(L-J) \times J$ polynomial matrix. Then 
 $({\bf U}_i^o(z^{-1}))^T$ is a left inverse of ${\bf V}_i^o(z)$. 
%
Conversely, any left inverse of ${\bf V}^o_i(z)$ has this form \cite{alan2}. 
A demultiplexing waveform
$w[n]$ is obtained by interleaving polynomials in  matrices
${\bf U}^o_i(z)$ in the same manner polynomials in 
${\bf V}^o_i(z),~i=0,1,\ldots, N/J$ are interleaved to obtain $v[n]$.
Modulated versions $\psi_k[n]$ of $w[n]$ in (\ref{eq:psi}) are then 
biorthogonal to waveforms $\varphi_k[n]$ according to (\ref{eq:biorth}). 
Polynomial matrices ${\bf A}_i(z)$ are free parameters which can be optimized
to achieve possible additional design requirements. Note that the biorthogonal
demultiplexing considered here is a generalized version of the cyclic prefix
or zero padding schemes and it is different from biorthogonal frequency division
multiplexing (BFDM), studied
 in \cite{kozek} and \cite{matz}, where multiplexing waveforms $\varphi_k[n]$ 
are not mutually orthogonal according to (\ref{eq:orth_fb1}). Some examples of
biorthogonal waveforms designed in this manner are presented in 
\cite{cvetkovic00}; below we illustrate the procedure for the case of 
rectangular waveforms $v[n]$.

\begin{example}
{
Consider the rectangular window for $N=128$ channel OFDM with $K=160$-sample 
frame interval, as discussed in Example \ref{ex:v128_160pp}.
Each of the corresponding matrices ${\bf V}^o_i(z)$ in
(\ref{eq:cpv}) can be completed to the square paraunitary matrix
\begin{equation}
{\bf V}^s_i(z)= {1 \over \sqrt{128}}\left[
\begin{array}{cc}
                   {\bf 0 }& 1 \\
                   {\bf I} & {\bf 0}^T
\end{array}
\right],~~~i=0,1,\ldots 31~,
\end{equation}
where ${\bf 0}=[0~0~0~0]$ and ${\bf I}$ is the $4 \times 4$ identity matrix. 
The corresponding matrices ${\bf U}^o_i(z)$ have the form
\begin{equation}
{\bf U}^o_i(z)= {1 \over \sqrt{128}}\left[
\begin{array}{cccc}
                   A_{i,1,1}(z) & A_{i,1,2}(z) & A_{i,1,3}(z)
                   & A_{i,1,4}(z) \\
                   1 & 0 & 0 & 0 \\
                   0 & 1 & 0 & 0 \\
                   0 & 0 & 1 & 0 \\
                   0 & 0 & 0 & 1
\end{array}
\right],~~~i=0,1,\ldots 31.
\end{equation}
The $K$ samples long rectangular demultiplexing window that is used
in the zero-padding scheme is obtained for
$A_{i,1,1}(z)=z$, $ A_{i,1,2}(z) =A_{i,1,3}(z)
=A_{i,1,4}(z) =0$ for all $i=0,1,\ldots 31$.}
\hfill$\square$
\end{example}

\subsection{Equivalent Orthogonality Conditions and the Relationship with Weyl-Heisenberg Frames}

The parameterization orthonormal Weyl-Heisenberg sets, presented in the previous subsection, allows for  exact orthogonal design using
unconstrained optimization procedures. Alternatively, if one wishes to pursue design using optimization under orthogonality
constraints in (\ref{eq:orth_fb1}), it is beneficial to find a minimal equivalent set of constraints expressed directly in terms 
of underlying prototype waveforms. To this end observe that
the orthogonality conditions
$\sum_n \varphi^\ast_m[n-kK]\varphi_l[n-iK]=\delta[m-l]\delta[k-i]$
can be written in terms of the prototype waveform, assuming it is real, 
as
$
e^{j{2\pi \over N}(mk-li)K}\sum_{n=-\infty}^{\infty}
v[n-kK]v[n-ik]e^{j {2 \pi \over N}(l-m)n} = \delta[m-l] \delta[k-i]~.
$
This is equivalent to
\begin{equation}
\sum_{n=0}^{N-1}V(i,k,n)e^{j {2 \pi \over N}(l-m)n}=\delta[m-l]
\delta[k-i]~,
\label{eq:der1}
\end{equation}
where
$
\displaystyle{V(i,k,n)=\sum_{p=-\infty}^{\infty} v[n-iK+pN]v[n-kK+pN].}
$
Note further that  (\ref{eq:der1}) is equivalent to
$
V(i,k,n)=\delta[k-i]/N,~n=0,1,\ldots, N-1,
$
which is satisfied if and only if
\begin{equation}
\sum_i v[n+iN]v[n+iN+jK]={1 \over N} \delta[j],~n=0,1,\ldots N-1~.
\label{eq:wconstraints}
\end{equation}
Hence, the orthogonality relations in (\ref{eq:orth_fb1}) are equivalent to  the conditions in (\ref{eq:wconstraints})
expressed directly in terms of $v[n]$.

One class of OFDM windows that follows immediately from 
(\ref{eq:wconstraints}) are windows which extend over only one 
frame interval, $K$, for   $K \leq 2N$.
With this restriction the system of constraints in
(\ref{eq:wconstraints}) reduces to
\begin{equation}
\begin{array}{ll}
v[n]v[n]+v[n+N]v[n+N]={1 \over {N}}, & 0 \leq n \leq K-N-1 \\
v[n]v[n]={1 \over {N}} & K-N \leq n \leq  N-1
\end{array},
\end{equation}
and the complete set of solutions  can be parameterized 
in terms of $K-N$ angles $\alpha_n$ as
\begin{equation}
v[n]={1 \over \sqrt{N}} \left\{ \begin{array}{ll}
              \cos(\alpha_n), & 0 \leq n \leq K-N-1 \\
              1,             & K-N \leq n \leq N-1 \\
              \sin(\alpha_{n-N}), & N \leq n \leq K-1
\end{array}
\right. .
\label{eq:param_short}
\end{equation}

The set of constraints  in (\ref{eq:wconstraints})  is also identical  to a set of sufficient and necessary  conditions under which 
the set
$\{\xi_{k,i}:\xi_{k,i}[n]=v[n-iN]e^{j {2 \pi \over
K}k(n-iN)}\}_{k \in {\Z}_K, i \in \Z}$ forms a tight frame in $\ell^2(\Z)$ \cite{cvetkovic00_1}.  Hence, the following theorem holds.
\begin{theorem} \label{th:tfexp}
{\em A Weyl-Heisenberg set
$\Phi=\{\varphi_{k,i}:\varphi_{k,i}[n]=v[n-iK]e^{j {2 \pi \over N}k(n-iK)}\}_{k \in \Z_N, i \in \Z}$ 
is orthonormal if and only if
$\Xi=\{\xi_{k,i}:\xi_{k,i}[n]=v[n-iN]e^{j {2 \pi \over K}k(n-iN)}\}_{k \in {\Z}_K, i \in \Z}$
 is a  tight frame in $\ell^2(\Z)$.}
\hfill$\square$
\end{theorem}
The result of Theorem \ref{th:tfexp}  has been
established previously for continuous-time Weyl-Heisenberg frames in 
\cite{janssen} and \cite{daubechies} and also pointed out in the context of OFDM in \cite{cvetkovic99,bolcskei99}. It can be proved along the lines
of  continuous-time proof, but that involves fairly sophisticated functional 
analysis and operator algebra. The proof provided here, on the other hand,
draws only upon elementary results on filter banks. 

Recently it was pointed out by Han and Zhang \cite{han} that when the transmitted sequence $a[n]$ takes values from a finite alphabet, linear independence of waveforms $\varphi_{k,i}[n]$ is not necessary for perfect demultiplexing. Furthermore, 
Han and Zhang propose using  Weyl-Heisenberg frames $\xi_{k,i}[n]$ instead of orthonormal families  $\varphi_{k,i}[n]$ 
 for transmission over time-frequency dispersive channels
and demonstrate some merits of this approach. A complete parameterization of  discrete-time  tight Weyl-Heisenberg frames was given in \cite{cvetkovic98}.

\section{OFDM Window Design Issues}

An  issue which is still a matter of debate
is whether the exact (bi)orthogonality
according to (\ref{eq:orth_fb1}) and (\ref{eq:biorth}), as pursued in 
this paper and in \cite{hleiss,kozek,strohmer,siclet,matz}, is really needed
considering that it is lost due to multipath propagation, frequency offset,
or timing mismatch, all of which occur simultaneously in a communication channel.
It is reasonable to think that it would be sufficient
to impose (bi)orthogonality of transmit and receive
waveforms only at neighbouring locations in the 
time-frequency lattice, {\sl i.e.}
\begin{equation}
\langle \psi_k[n-iK] ,\varphi_l[n-jK]\rangle=\delta[k-l]\delta[i-j],~
|k-l|<N_0,~|i-j|<K_0~,
\label{eq:porth}
\end{equation}
for some small $N_0$ and $K_0$, 
and use the design
freedom acquired by relaxing the remaining constraints  
to achieve higher spectral containment of the waveforms. This
better frequency localization would in turn 
provide near-orthogonality at other lattice points and  
also improve robustness to the considered sources of degradation.
Such an approach was considered in \cite{holte96,hass}.
A more radical strategy would be to abandon the orthogonality 
completely and 
suppress interchannel interference by further maximizing spectral containment 
of the
waveforms, as proposed in \cite{cherubini,cherubini02}, or  minimize ISI/ICI 
explicitly if the channel or its statistics are known
\cite{lin,phoong}. 
Simulation results which compare a scheme with a partial orthogonality
as specified in (\ref{eq:porth}) to a scheme with the orthogonality across
the whole time-frequency lattice reported in \cite{kozek}
showed a superior performance of the latter scheme. In the early
phase of this research, we also considered a variant of the partial orthogonality
approach, in particular $\langle \varphi_k[n-iK], \varphi_l[n-jK]\rangle=
\delta[i-j]$ for $l=k$ only, but the results were inferior compared to
the case of full orthogonality despite
better frequency localization. 
Recently, Matz {\sl et al.} showed in the continuous time case that optimal waveforms for transmission over
time-frequency dispersive channels are very close to orthogonal \cite{matz}.
In this section we provide some additional insight
into why the exact orthogonality has merits in reducing the interference.
Further, we show that, contrary to what is believed by many practitioners,
tapering of the rectangular demultiplexing window in the zero-padding (or
cyclic prefix)
scheme cannot improve the robustness of OFDM to frequency offset.

Consider transmission of an OFDM signal through a multipath
channel. The received signal $s_r[n]$ has  the form
$
s_r[n]=\sum_{l=0}^D r_l s[n-d_l]
$
where $s[n]$ is the transmitted signal
as given by (\ref{eq:dtmodel1}).
Assume without loss of generality that  $r_0=1$, $d_0=0$, and assume further
that 
the channel does not change with time.
Sequences $\tilde{a}_{k}[i]$ at the output of the
demultiplexer,
$\displaystyle{\tilde{a}_{k}[i]=\sum_n \phi_k[iK-n]s_r[n],}$
have the form
$
\tilde{a}_{k}[i]=a_{k}[i]F_{k,i}+\sum_{(p,q)\neq (0,0)}a_{k+p}[i+q]
I_{k,i}(p,q),
$
where $F_{k,i}$ and $I_{k,i}(p,q)$ are fading and intersymbol interference
functions, respectively.  These two functions can be expressed in
terms of the crossambiguity function between the multiplexing and 
demultiplexing modulating waveforms,
$
\displaystyle{{\cal A}_{v,w}(x,y)=\sum_n v[n-x]w[-n] e^{j {2 \pi \over N}yn},}
$
as
$
F_{k,i}= 1+\sum_{l=1}^D r_l e^{-j{2 \pi \over N}kd_l}{\cal A}_{v,w}(d_l,0),
$ and 
$
I_{k,i}(p,q)=e^{-j {2 \pi \over N}(k+p)qK} \left( {\cal A}_{v,w}(qK,p)+
\sum_{l=1}^{D}r_l e^{-j{2\pi \over N}(p+k)d_l}{\cal A}_{v,w}(qK+d_l,p)\right).
$
Assuming that  symbols $a_k[i]$ are independent,
zero-mean,
with  variance $1$, $E(a_k[i]a^\ast_l[j])=\delta[k-l]\delta[i-j],$
and that  channel  parameters $r_l$ are also
statistically independent, 
$E(r_lr_m^\ast)=\delta[l-m]\sigma_l^2,$
the expected squared value of the total interference 
$
\displaystyle{
{\cal I}=E(|\sum_{(p,q)\neq (0,0)} a_{k+p}[i+q]I_{k,i}(p,q)|^2)},$
is given by
$$
{\cal I}=\sum_{(p,q) \neq (0,0)}\left(|{\cal A}_{v,w}(qK,p)|^2+
\sum_{l=1}^D\sigma_l^2 |{\cal A}_{v,w}(qK+d_l,p)|^2\right).
$$
Design methods which
minimize this interference without imposing biorthogonality
between multiplexing and demultiplexing waveforms 
and assuming that there is no frequency offset nor timing 
mismatch and statistics of the channel are known were studied in detail in
\cite{scaglione1,scaglione2} for waveforms up to $K$ samples long, and more recently for
longer waveforms in 
\cite{phoong}.  
When an OFDM system operates 
with a frequency offset $\epsilon_f$ and a timing mismatch of $\epsilon_t$
samples the expected squared value of  intersymbol interference becomes
$$
{\cal I}=\sum_{(p,q) \neq (0,0)}\left(|{\cal A}_{v,w}(qK-\epsilon_t,p+
\epsilon_f)|^2+\sum_{l=1}^D\sigma_l^2 |{\cal
A}_{v,w}(qK+d_l-\epsilon_t,p+
\epsilon_f)|^2\right).
$$
To suppress
this interference the waveforms should be designed so
that ${\cal A}_{p,q}(x,y)$ has low magnitude
in all regions ${\cal E}_{p,q}=(pK-\epsilon_t^-,
pK+\Delta+\epsilon_t^+) \times (q-\epsilon_f^-,q+\epsilon_f^+)$
where  $(p,q)$ is a pair of integers different from $(0,0)$, 
$\Delta$ is the maximal
delay spread, and  $\epsilon_t^-$, $\epsilon_t^+$, $\epsilon_f^-$ and
$\epsilon_f^+$ are bounds on possible timing mismatch and
frequency offset.
Ideally, ${\cal A}_{v,w}(x,y)$ should vanish in all regions of this
form. However,  with finite-length waveforms ${\cal A}_{v,w}(x,y)$ can
have only a limited number of zeros within each region ${\cal E}_{m,n}$.
For example, given a multiplexing waveform $v[n]$ and assuming that 
$v[n]$ and $w[n]$ are of the same length, $L_v$,
the requirement that ${\cal A}_{v,w}(x,y)$ has $n_z$ zeros within each
${\cal E}_{m,n}$ imposes
$n_z\left(2NL_v/K -N-1\right)$
constraints on $w[n]$. It follows that if the 
efficiency $N/K$ is to be above $50\%$, and the waveforms are at least
$2N$ samples long,  $n_z$ cannot be more than $1$, except in
singular cases such as the zero-padding or cyclic prefix schemes.
Hence, designing $v[n]$ and $w[n]$ so that: (i)  ${\cal A}_{v,w}$ has a zero in 
each of the regions ${\cal E}_{m,n}$, {\sl e.g.}
\begin{equation}
{\cal A}_{v,w}(pK,q)=0,~{\rm for ~all~integer~pairs~}(p,q) \neq (0,0)~,
\label{eq:v(xy)orth}
\end{equation}
and (ii) they do not have  sharp transitions, or have a somewhat flat
shape within the main lobe in time, to avoid rapid changes of the crossambiguity 
function around its zeros,
seems to be a good window design strategy. Note that the condition in
(\ref{eq:v(xy)orth}) is equivalent to the
biorthogonality between the multiplexing and  demultiplexing
waveforms  in (\ref{eq:biorth}).
This suggests that although it is not clear whether this strict
biorthogonality is crucial for the effectiveness  of OFDM,
it is certainly a judicious design approach.
In addition to the smoothness and the orthogonality requirements,
in order  to minimize interference between different OFDM
channels and make the system robust to frequency offset, the waveforms
should have adequate frequency characteristics, which basically means
good spectral containment in the $[0,{\pi \over N}]$ band. 
One way of balancing all these requirements is by designing the waveforms to
optimize their time-frequency localization as 
concluded also in \cite{matz} for the continuous-time case.

In the case of the cyclic prefix or zero padding schemes the  crossambiguity 
function ${\cal A}_{v,w}(x,y)$ is
zero at all integer pairs $(x,y)$ such that $mK \leq x \leq mK+T_g$.
Note that this is a singular case where
the modulating waveforms are constant within their support, and this 
time-invariance   enables having the crossambiguity 
function with $K-N$
zeros in all  regions ${\cal E}_{m,n}$. However, ${\cal A}_{v,w}$ grows
rapidly around its zeros and that makes these
schemes very sensitive to timing mismatch and frequency
offset. The problem of timing mismatch can be dealt with by
introducing an adequate time-shift in the demultiplexing waveform
at the expense of a reduction in the tolerable delay spread.
The problem with the sensitivity to frequency offset, on the other hand, cannot be
approached in such a straightforward manner.
It might appear that using a demultiplexing waveform which is constant on
an interval of length $N+d_s$ and has a gradual decay toward zero
would improve the robustness of 
these  schemes to frequency offset owing to a better 
frequency
localization of the waveform, while intersymbol
interference would be still
eliminated in the absence of frequency offset for delay spreads up to
$d_s$ samples. To demonstrate that this kind of  modification cannot
improve the robustness to frequency
offset, consider a demultiplexing waveform $w_r[n]$ given by
\begin{equation}
w_r[n]=\left\{ \begin{array}{ll}
              r_1[n], & -K+\alpha+1 \leq n \leq -N-d_s \\
              {1 \over \sqrt{N}} & -N-d_s+1 \leq n \leq 0 \\
              r_2[n] & 1 \leq n \leq \alpha
             \end{array} \right.
\end{equation}
where $r_1[n]$ and $r_2[n]$ are roll-off functions with support
of $\alpha=(K-N-d_s)/2$ samples, so that the total duration of $w_r[n]$ is
$K$ samples. The crossambiguity function, ${\cal A}_{v,w_r}$, between this
waveform and the $N$-samples long rectangular waveform
is ${\cal A}_{v,w_r}(x,y)={1 \over N} \sum_{n=x}^{n=x+N}e^{j{2\pi\over
N}yn}$ for $x$ in the range $0 \leq x \leq d_s$, and this is identical to the
crossambiguity function obtained with the $K$-samples long rectangular
demultiplexing waveform. Hence, this
modification, even though it provides a better spectral containment of the
demultiplexing waveform does not improve the robustness of the zero-padding 
scheme to frequency offset.

\section{Design Examples and Experimental Results}
\label{sec:experiments}

\begin{figure}[thb]
\begin{tabular}{ccc}
\includegraphics[width=0.45\textwidth]{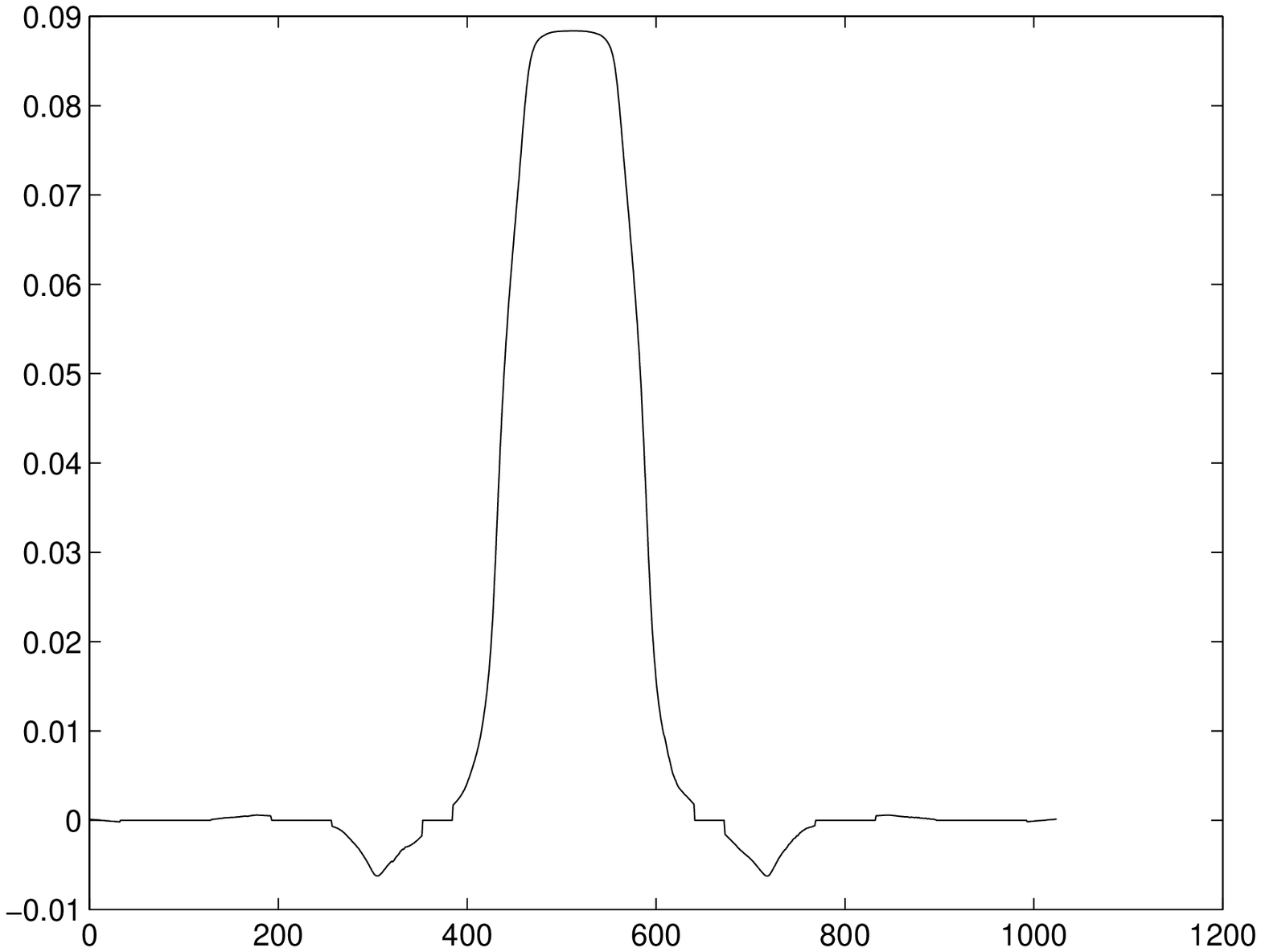} & \hspace*{0.5cm} &
\includegraphics[width=0.45\textwidth]{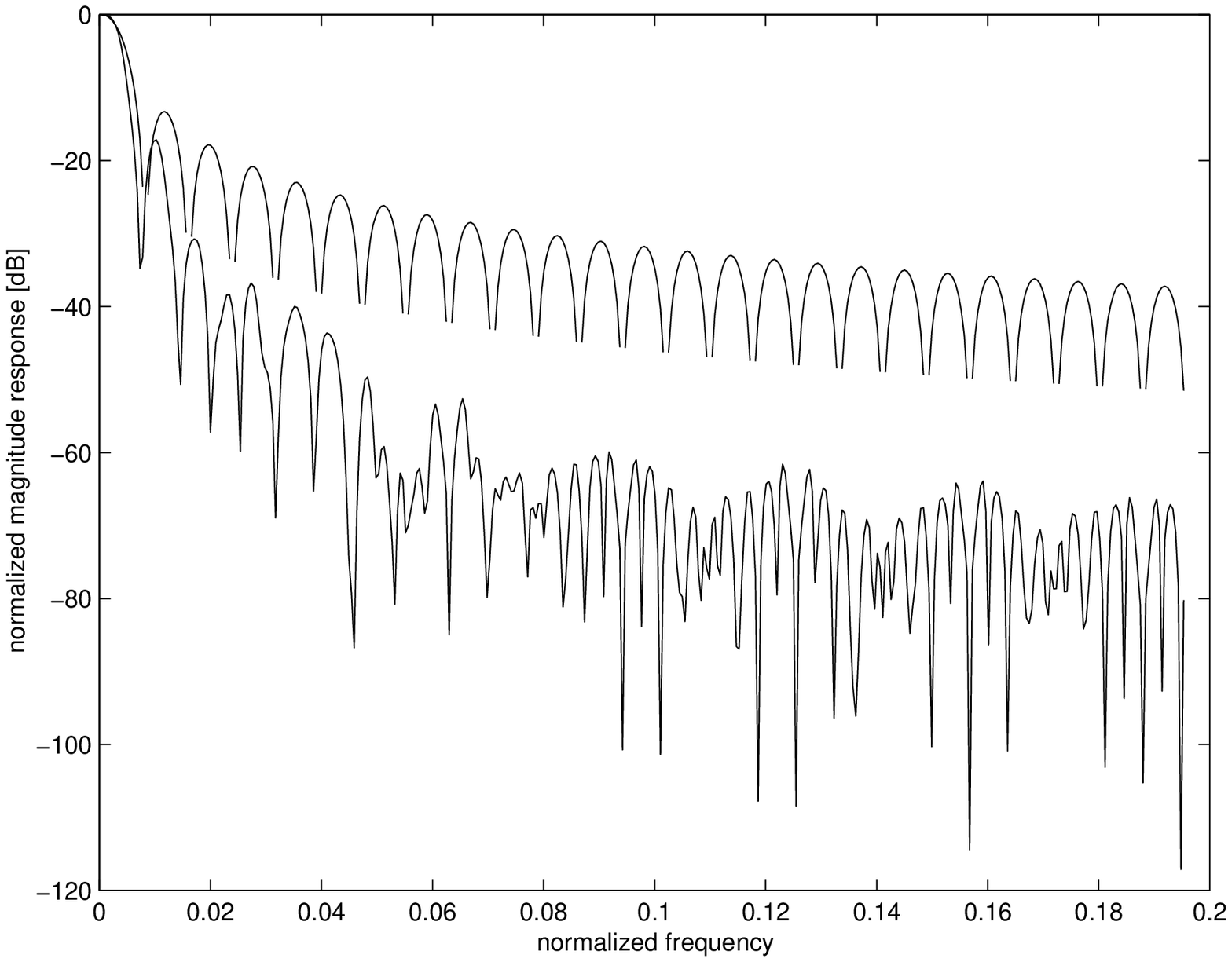} \\
a) & & b)\\
\includegraphics[width=0.45\textwidth]{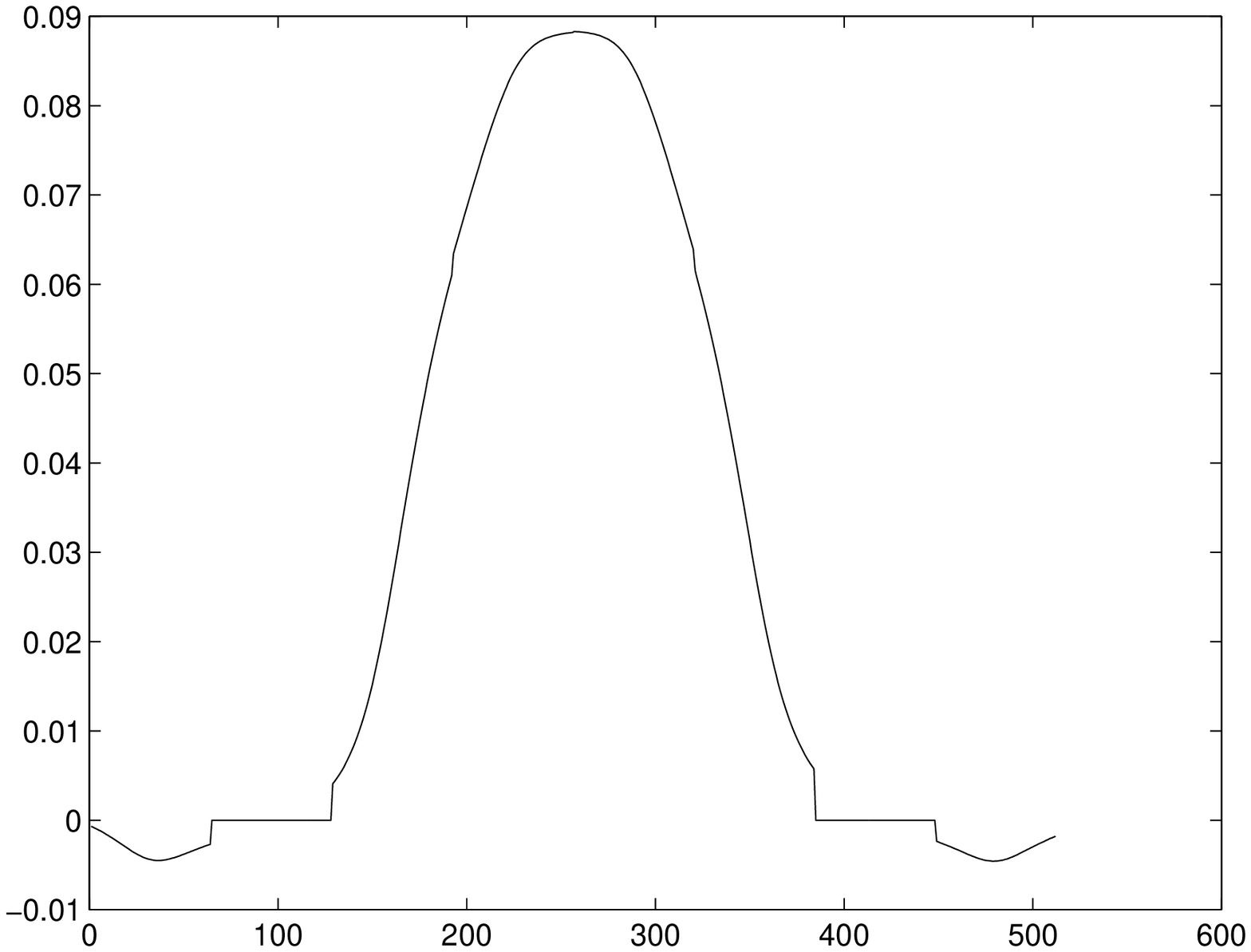}
 & \hspace*{0.5cm} &
\includegraphics[width=0.45\textwidth]{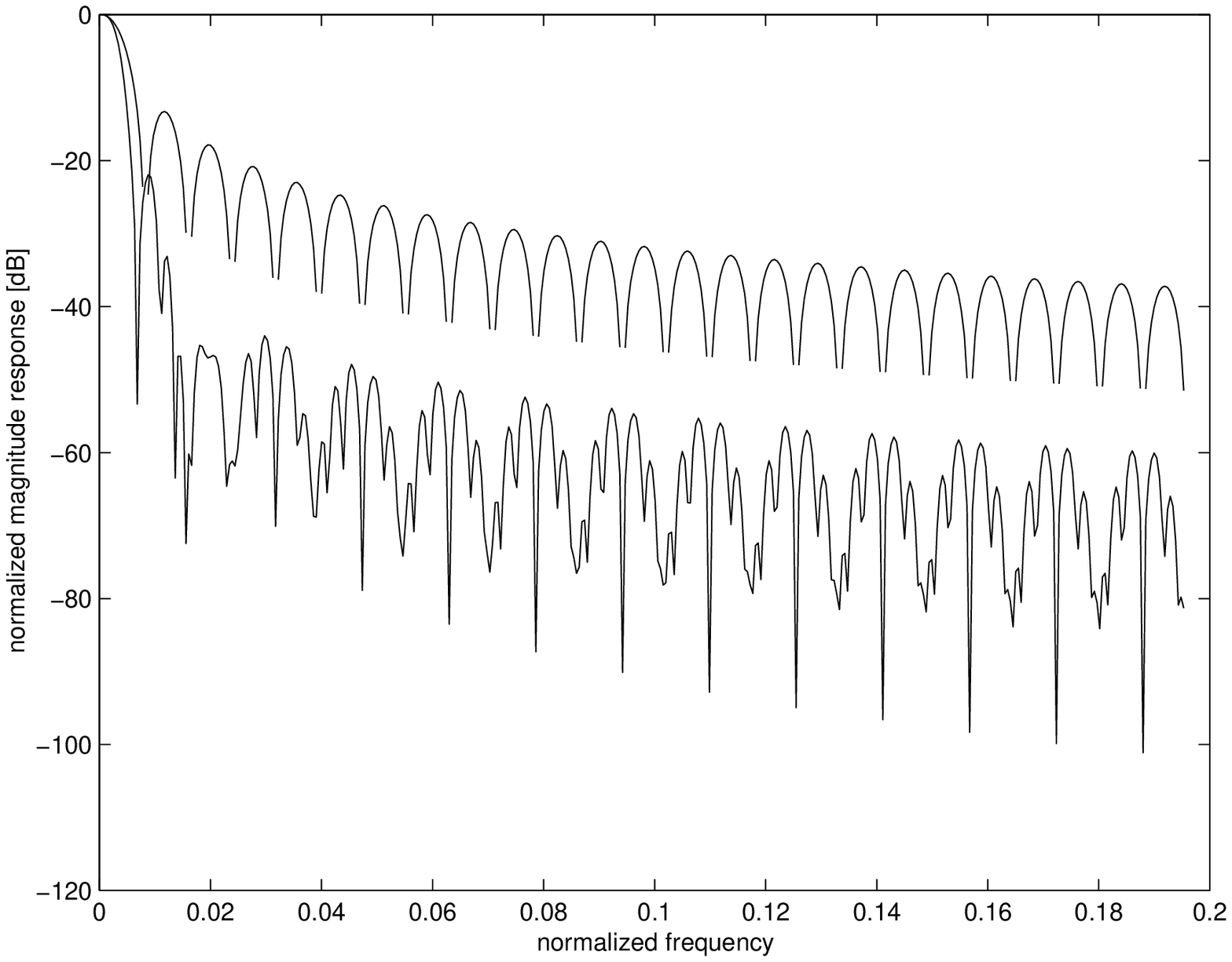} \\
c) & & d) \\
\end{tabular}
\caption{\sl Modulating waveforms for  for $N=128$-channel OFDM. 
a) Waveform $v_{160}$ for OFDM
with $K=160$-sample frame interval. b)
Magnitude responses of $v_{160}$ and the $128$-tap rectangular window.
c) Waveform $v_{192}$ for OFDM
with $K=192$-sample frame interval. d)
Magnitude responses of $v_{192}$ and the $128$-tap rectangular window.
}
\label{fig:w160_1024}
\end{figure}

In this section we present several examples of OFDM modulating waveforms.
First, we present examples of windows  designed for $N=128$ channel OFDM and
frame intervals of $K=160$ and $K=192$ samples.
The waveforms are designed by minimizing a convex combination of their
energy leakage out of the $[0,\pi/N]$ frequency band and their energy
leakage outside of the main lobe in time. The minimization is performed as
an unconstrained optimization using the parameterizations given in
Section \ref{sec:parameterization} combined with parameterizations of
paraunitary matrices based on Given's rotations \cite{ppv,martin}.  
Other optimization criteria can also be used
as discussed in \cite{siohan,siclet,matz}.
Waveform $v_{160}$ for $N=128$ channel OFDM with
$K=160$-sample frame interval obtained in this manner, as discussed in Example \ref{ex:v128_160pp}, is shown in 
Figure \ref{fig:w160_1024}a). Waveform $v_{192}$ shown in Figure \ref{fig:w160_1024}c) is designed
for $128$ channel OFDM with $K=192$-sample frame interval.
%

\begin{figure}[thb]
\begin{tabular}{ccc}
\includegraphics[width=0.45\textwidth]{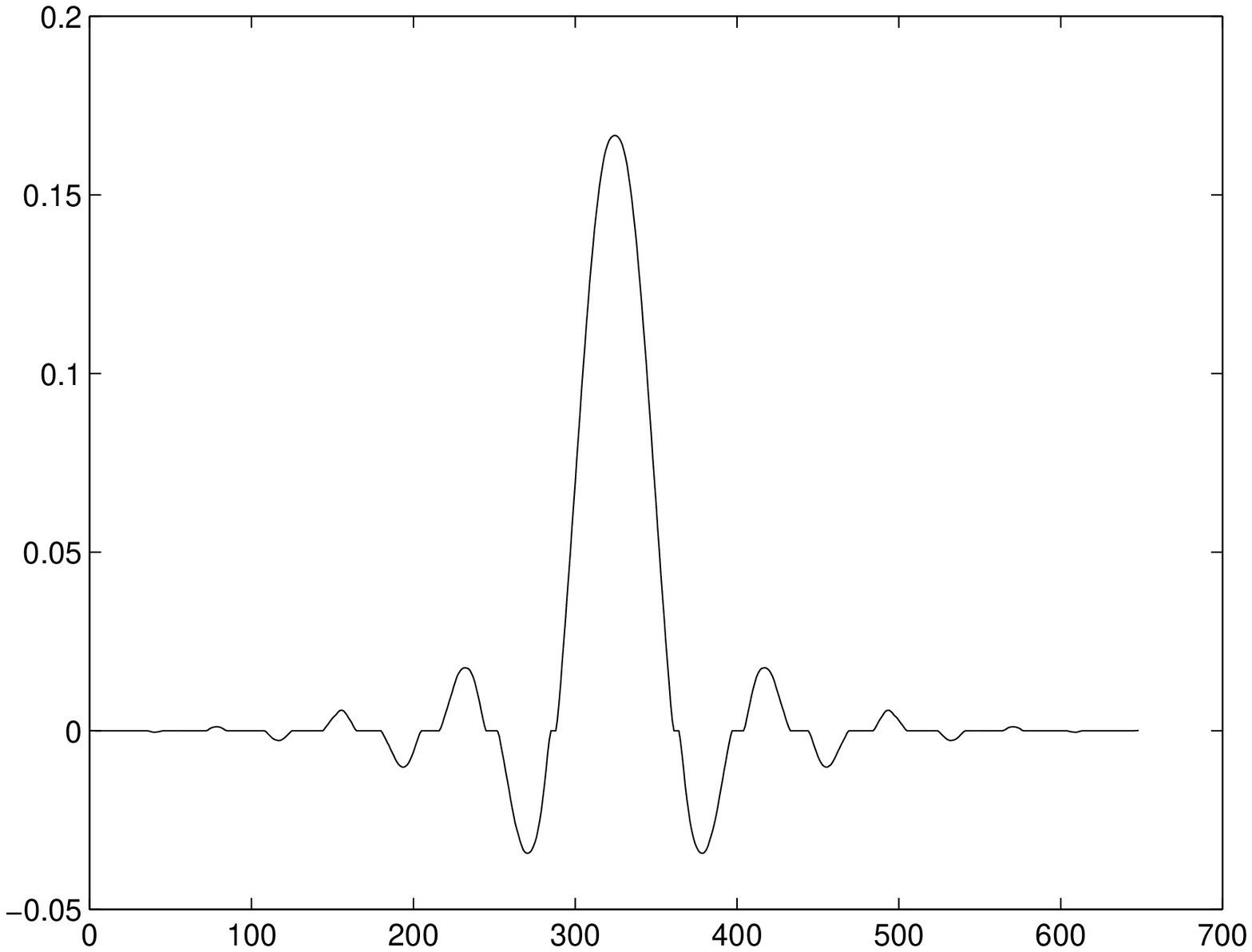} & \hspace*{0.5cm} &
\includegraphics[width=0.45\textwidth]{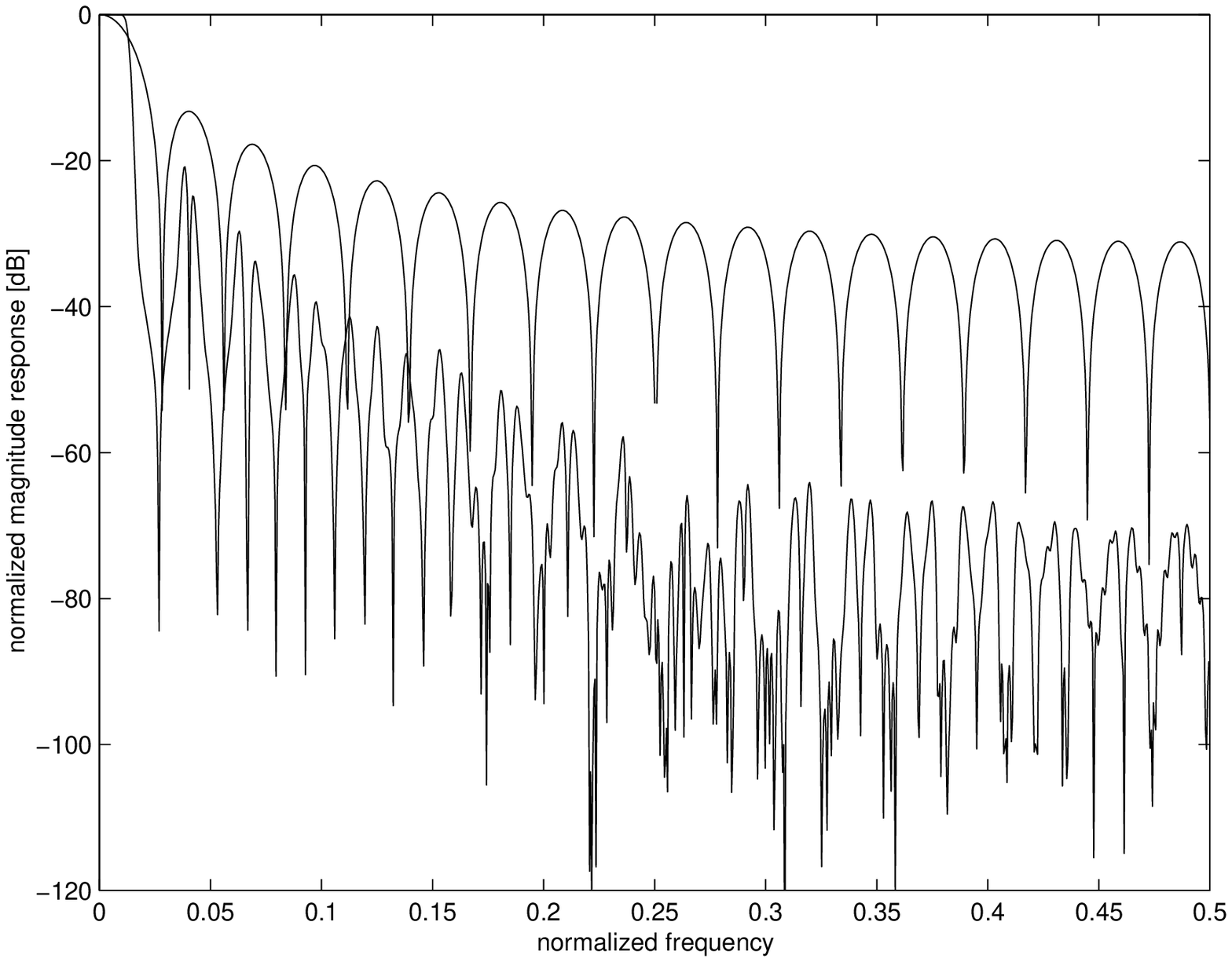} \\
 a) &  & b) \\
\includegraphics[width=0.45\textwidth]{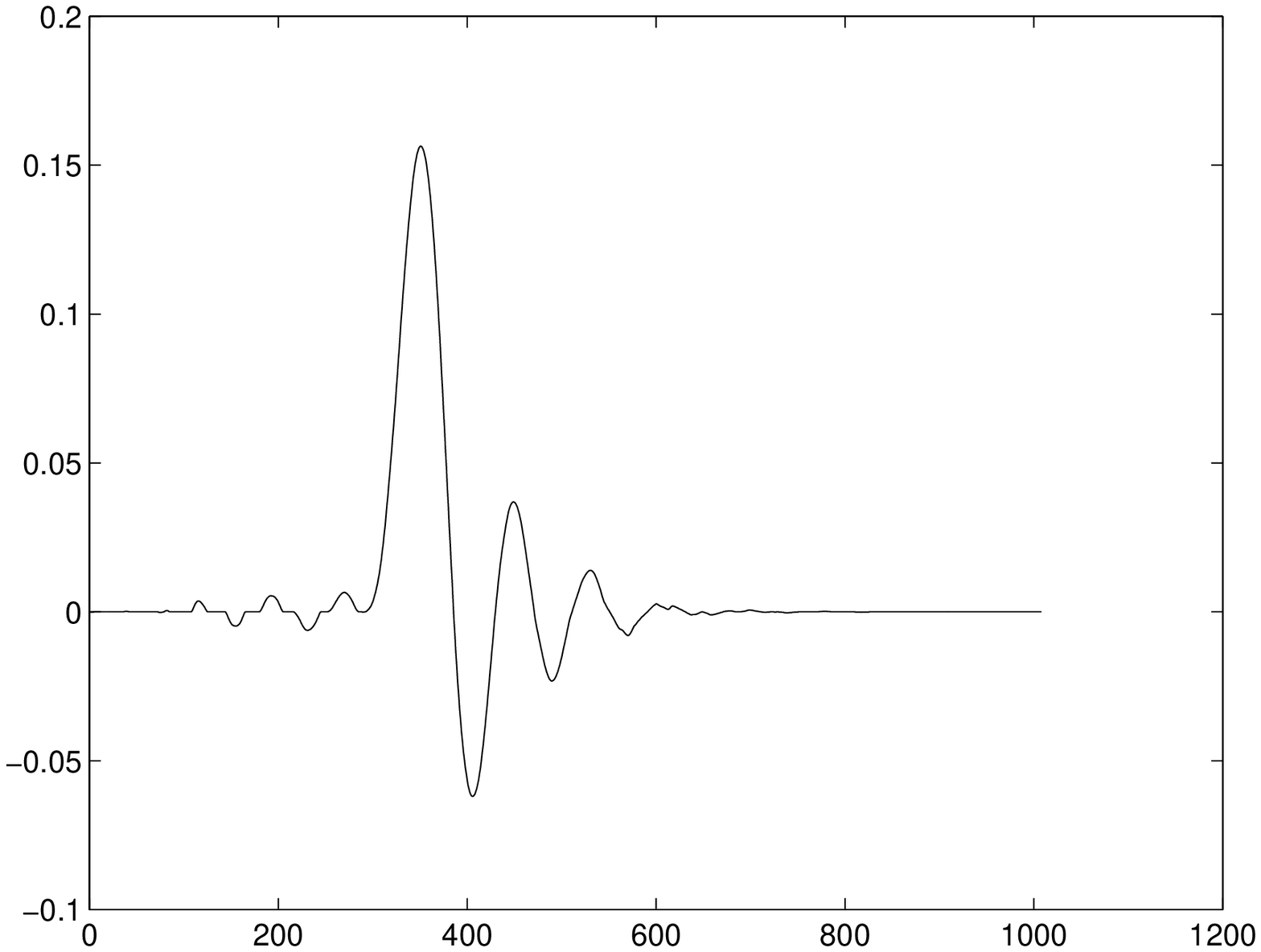} & \hspace*{0.5cm} &
\includegraphics[width=0.45\textwidth]{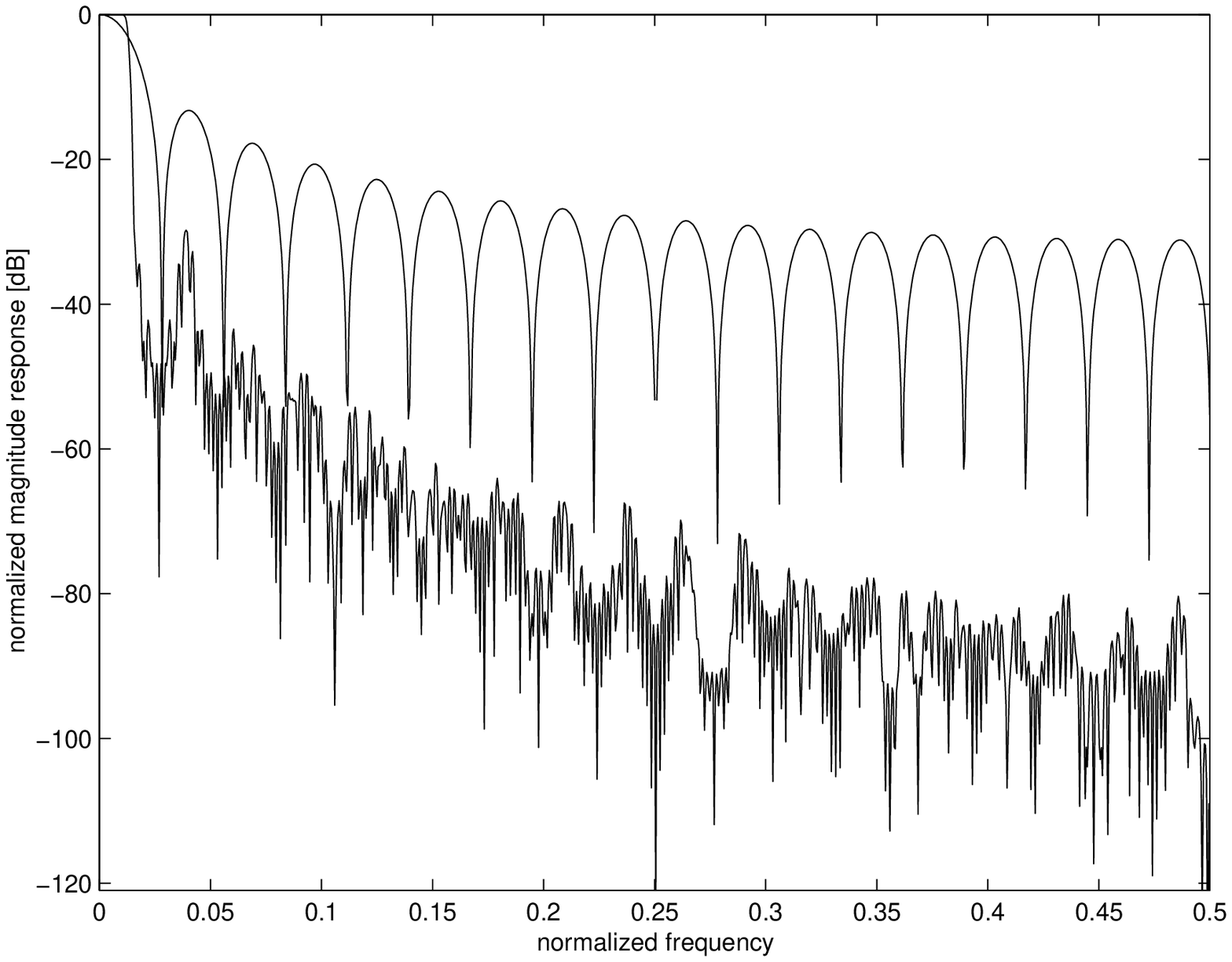} \\
 c) &  & d)
\end{tabular}
\caption{\sl Waveforms for $N=36$-channel OFDM with $K=40$-sample
frame interval. a) A symmetric waveform with $360$ nonzero
taps. b)  Magnitude responses of the symmetric waveform and 
the $36$-tap rectangular window. 
c) An asymmetric waveform with $720$ nonzero taps.
d) Magnitude responses of the asymmetric waveform and
the $36$-tap rectangular window. 
}
\label{fig:cable}
\end{figure}

Second, we present examples of  OFDM waveforms  for
upstream cable TV channels. Delay spread in cable TV channels is relatively
short compared to frame intervals, so it does not cause
considerable intersymbol interference. The major source of degradation
in these  channels is narrowband
interference, {\sl ingress}. The objective of OFDM modulation in such 
conditions is to enable
efficient transmission through the part of the spectrum which
is not affected by ingress. The main OFDM waveform design objective is
therefore to achieve a high spectral containment to limit the adverse 
effects of
ingress only to tones directly affected. The two waveforms shown in Figure
\ref{fig:cable} are designed for
transmission at $1440$kbaud ($36$ symbols per $25\mu {\rm s}$) through
a $1600$kHz cable channel, as specified in \cite{vcmt_prop}.
Both waveforms corresponds to $N=36$-channel OFDM with $K=40$-sample
frame interval. The waveform in Figure \ref{fig:cable}a) is symmetric
with $360$ nonzero taps, while the waveform in Figure \ref{fig:cable}c) is
asymmetric, with
$720$ nonzero taps, and hence lower sidelobes.

\begin{figure}[thb]
\centerline{
\begin{tabular}{ccc}
\includegraphics[width=0.45\textwidth]{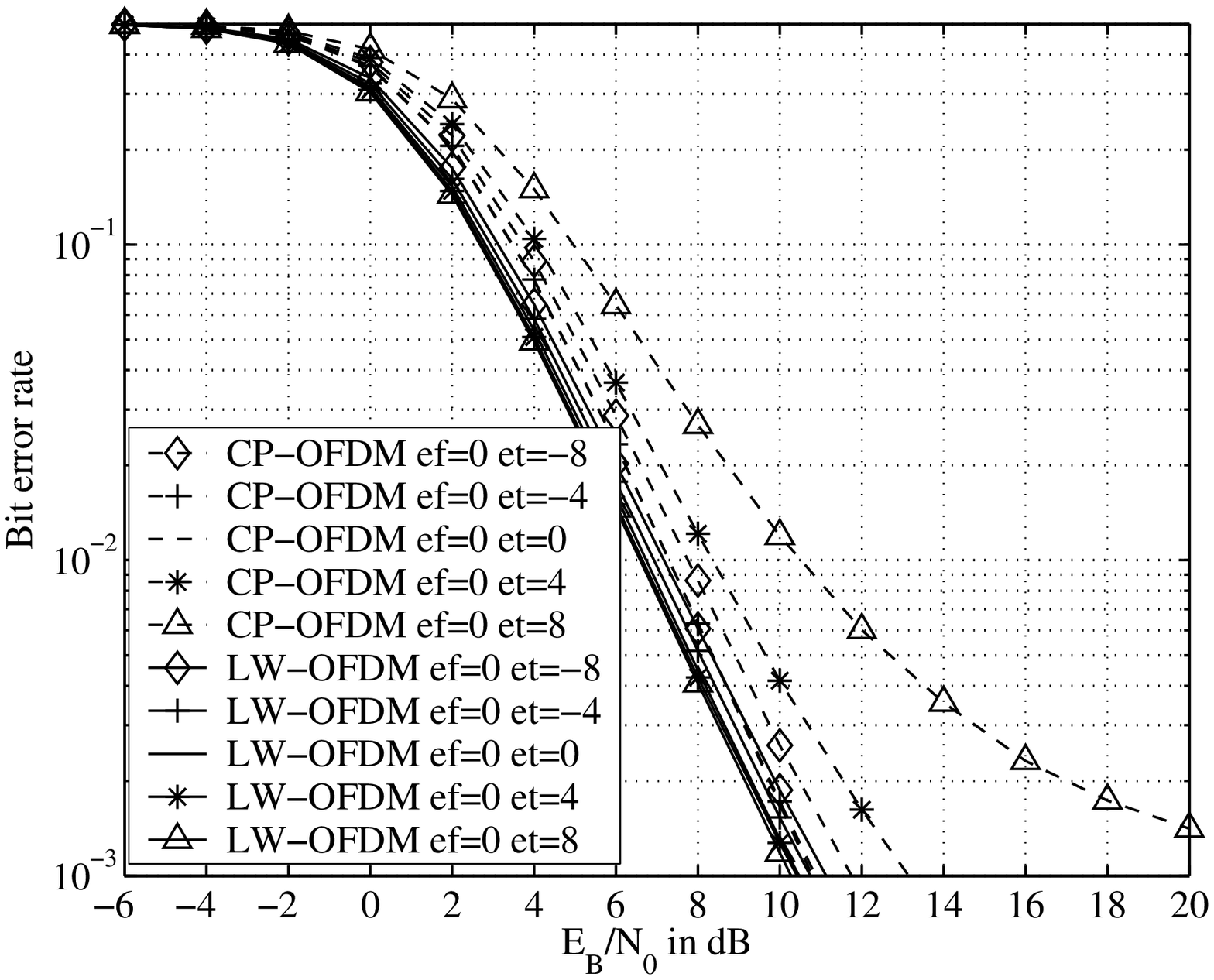}
 & \hspace*{0.5cm} &
\includegraphics[width=0.45\textwidth]{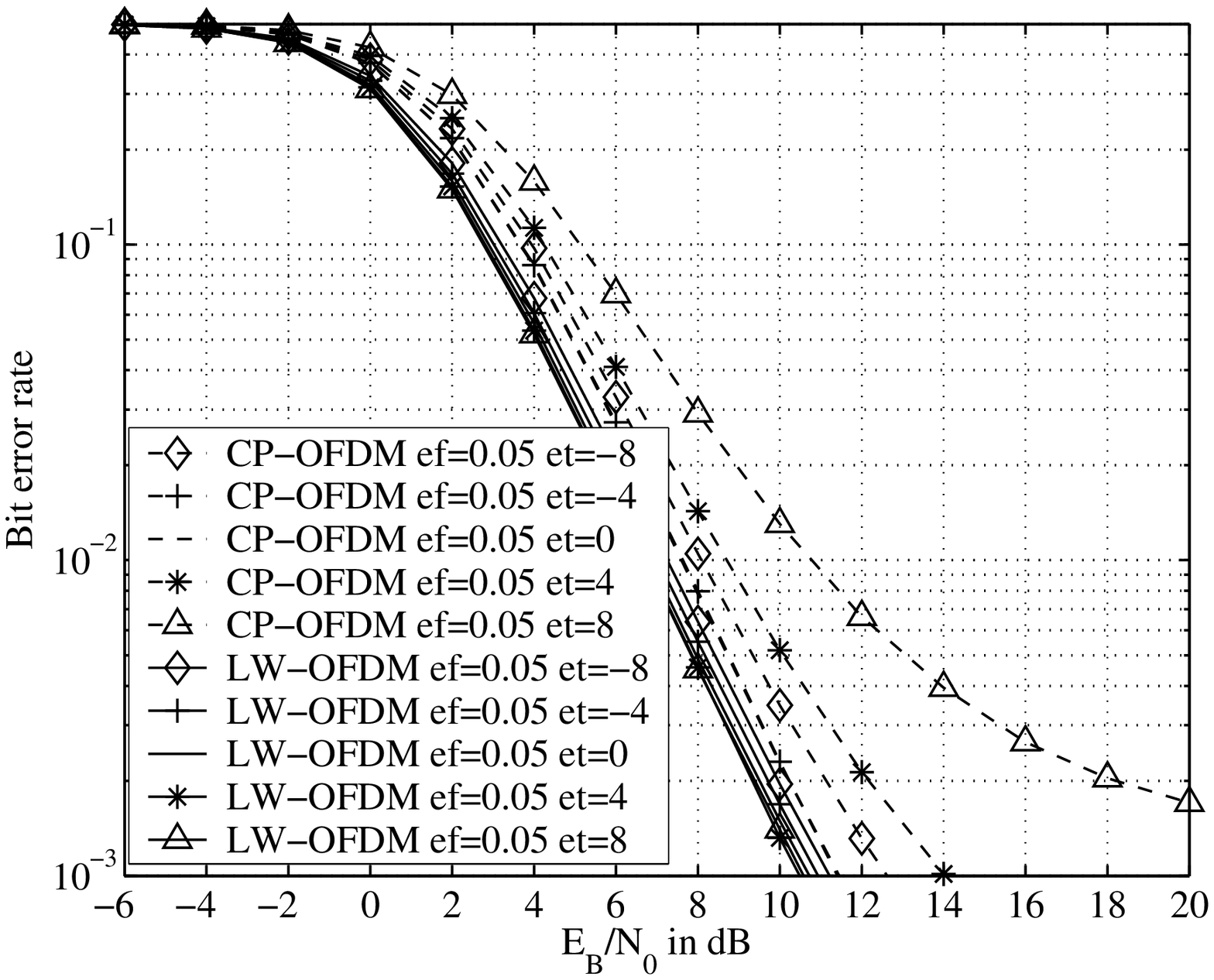} \\
a) & & b) \\
\includegraphics[width=0.45\textwidth]{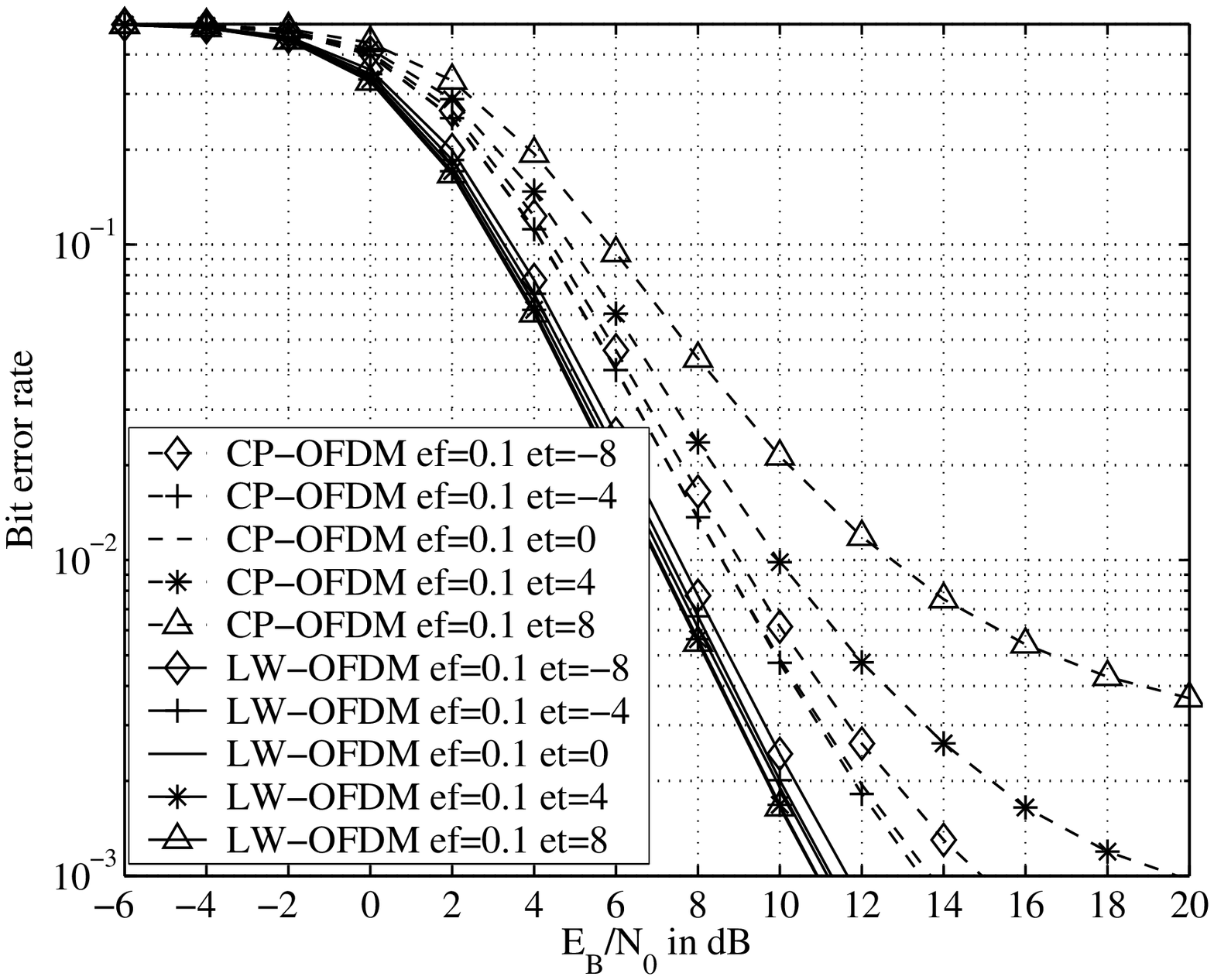}
 & \hspace*{0.5cm} &
\includegraphics[width=0.45\textwidth]{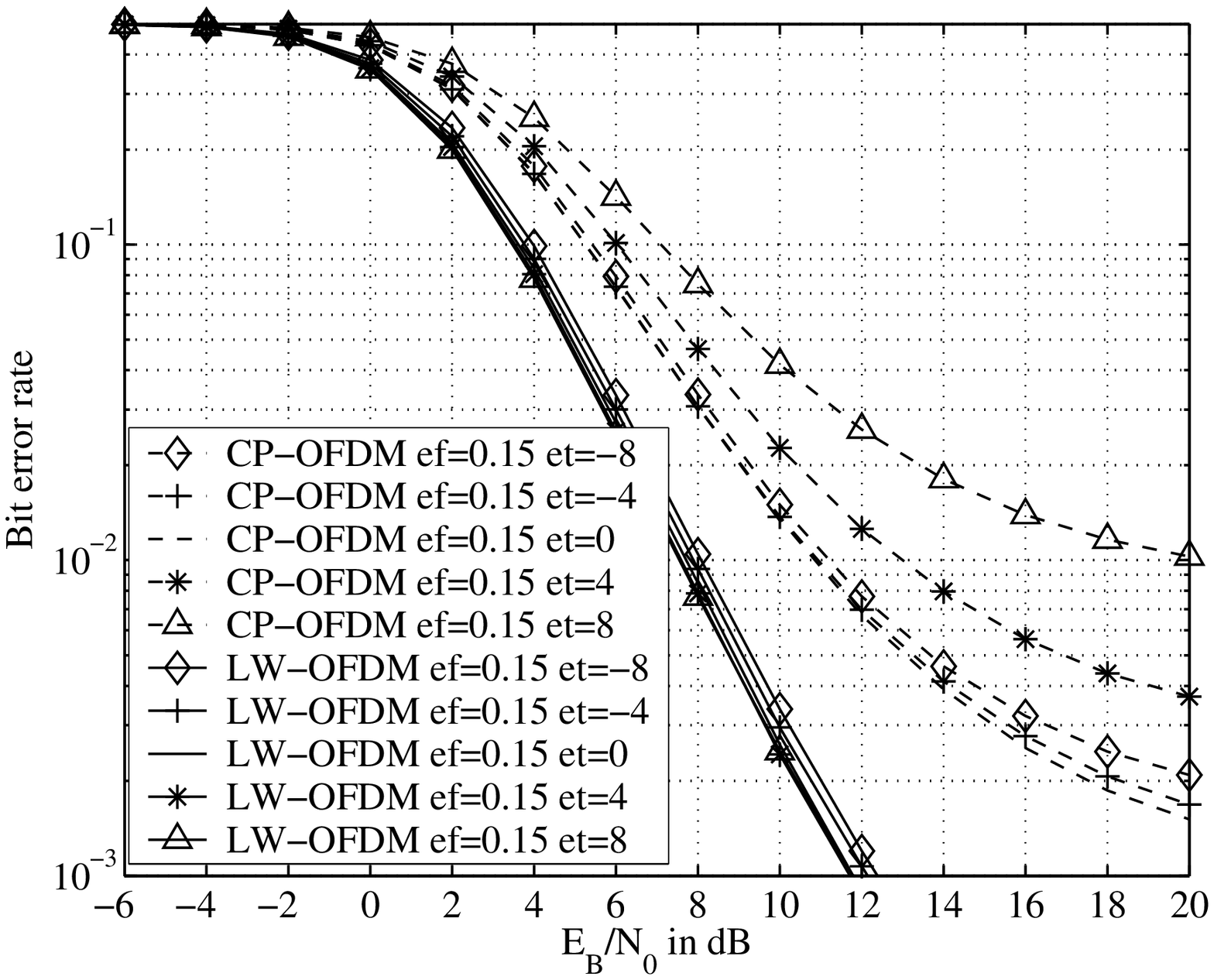} \\
c) & & d)
\end{tabular}}
\caption{\sl Measured bit error rates as a function of $E_{B}/N_{0}$
values for $N=128$-channel OFDM with $K=160$-sample frame interval
in transmissions over a $33$-tap Rayleigh channel in the presence of
frequency offset ($ef$) and timing mismatch ($et$). 
Two schemes are used: the cyclic prefix scheme  (CP-OFDM)
and OFDM based on waveform $v_{160}$ shown in Figure \ref{fig:w160_1024}a)
(LW-OFDM). 
Frequency offset
($ef$) and timing mismatch ($et$) are set to the values given in the
legend.  a) $ef=0$ - no frequency
offset.
b) $ef=0.05$ - frequency offset is $5\%$ of the bandwidth of one tone.
c) $ef=0.1$ - frequency offset is $10\%$ of the bandwidth of one tone.
b) $ef=0.15$ - frequency offset is $15\%$ of the bandwidth of one tone.}
\label{fig:ber160}
\end{figure}

We simulated OFDM based on long waveforms proposed in this paper in some 
typical transmission scenarios and compared the results with
cyclic prefix schemes which use same levels of redundancy. Orthogonal 
demultiplexing is used as opposed to biorthogonal demultiplexing in line
with the observation made in \cite{matz} that optimal performance is
obtained with close-to-orthogonal demultiplexing. 
Figures \ref{fig:ber160}a) to \ref{fig:ber160}d) show bit
error rates as a function of $E_{B}/N_{0}$ values 
(signal-to-noise ratio) for  $N=128$-channel OFDM with $K=160$-sample
frame interval in the presence of multipath propagation,
additive white Gaussian noise, frequency offset and timing mismatch.
The error rates were measured for OFDM based on waveform
$v_{160}$ shown in  Figure \ref{fig:w160_1024}a) and for the  
cyclic prefix scheme with $K-N=32$ samples long cyclic prefix. 
The simulated channel was a $L=33$-tap Rayleigh channel. 
The average powers of the taps in the ascending order were $0$, $-1$, $-2$,
$-3$, $-4$, $-5$, $-6$, $-7$, $-8$, $-9$, $-10$, $-11$, $-12$,
$-13$, $-14$, $-15$, $-16$, $-17$, $-18$, $-19$, $-20$, $-21$,
$-22$, $-23$, $-24$, $-1$, $-2$, $-3$, $-4$, $-5$, $-6$, $-7$, and
$-8$dB. The transmitted data symbols were encoded  using a rate $R=1/2$
convolutional  code  and  then BPSK
modulated. The one-tap equalizer was trained  using one training
frame, assuming the absence of noise during the transmission
of this frame. Frequency offset and timing mismatch were, on the other hand,
present during the training process.

The simulation results shown in Figure \ref{fig:ber160} were obtained
for frequency offsets $\epsilon_f = 0,$
$0.05,$ $0.10,$ $0.15$ ({\sl i.e.} 
up to the frequency offset equal to $15\%$ of the bandwidth of one tone),
and  timing mismatch of
$\epsilon_t =-8,-4,0,4,8$ samples. 
While the  error
rates of the cyclic prefix scheme (CP-OFDM)
increase significantly with  impairments, OFDM based on 
$v_{160}$ (LW-OFDM) exhibits robust behavior.
Some preliminary simulations, the results of which are not reported here,
indicate that the discrepancy between the robustness of the 
cyclic prefix scheme
and OFDM based on long waveforms  becomes more pronounced as the redundancy
increases.

\begin{figure}[thb]
\centerline{
\begin{tabular}{ccc}
\includegraphics[width=0.45\textwidth]{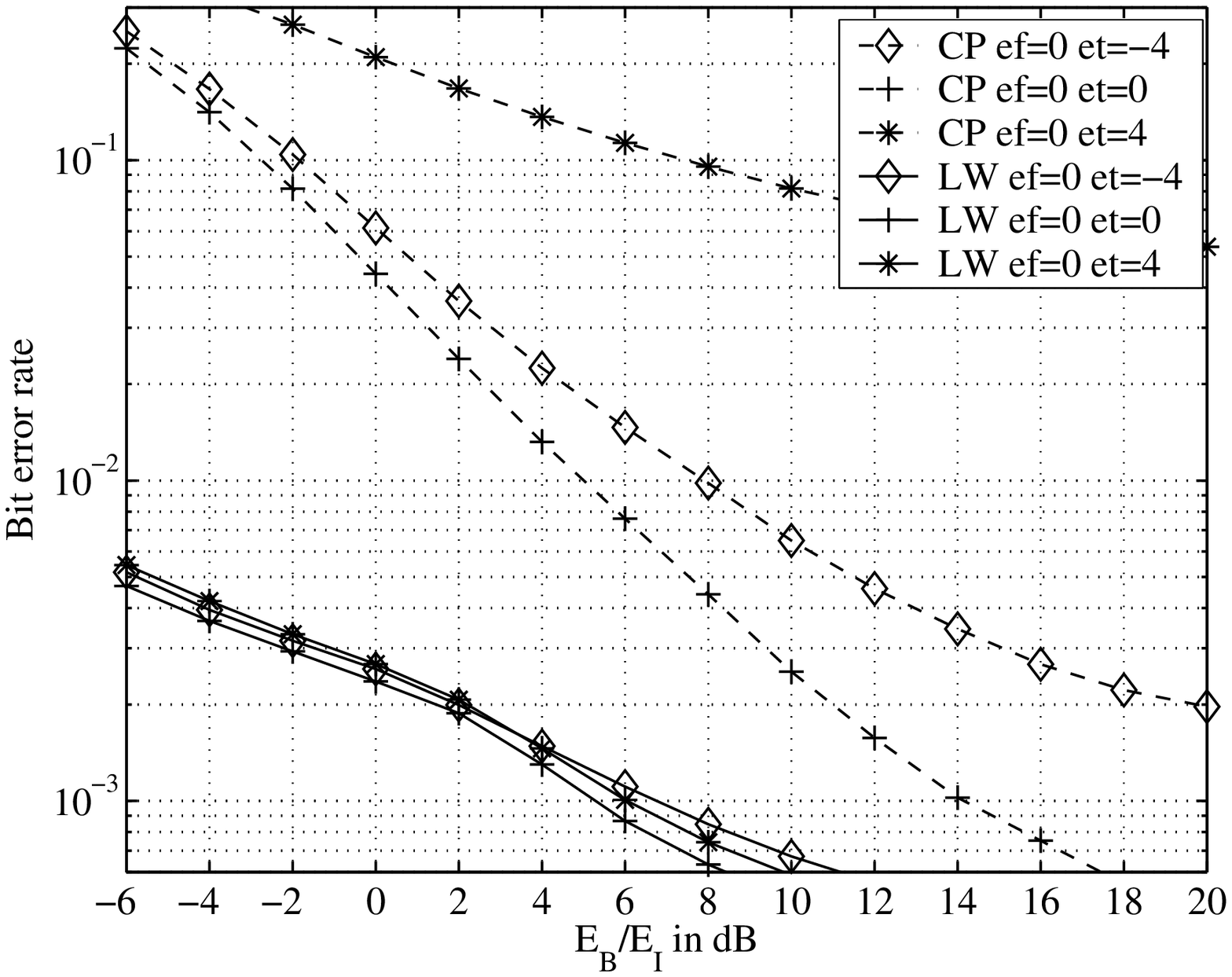}
 & \hspace*{0.5cm} &
\includegraphics[width=0.45\textwidth]{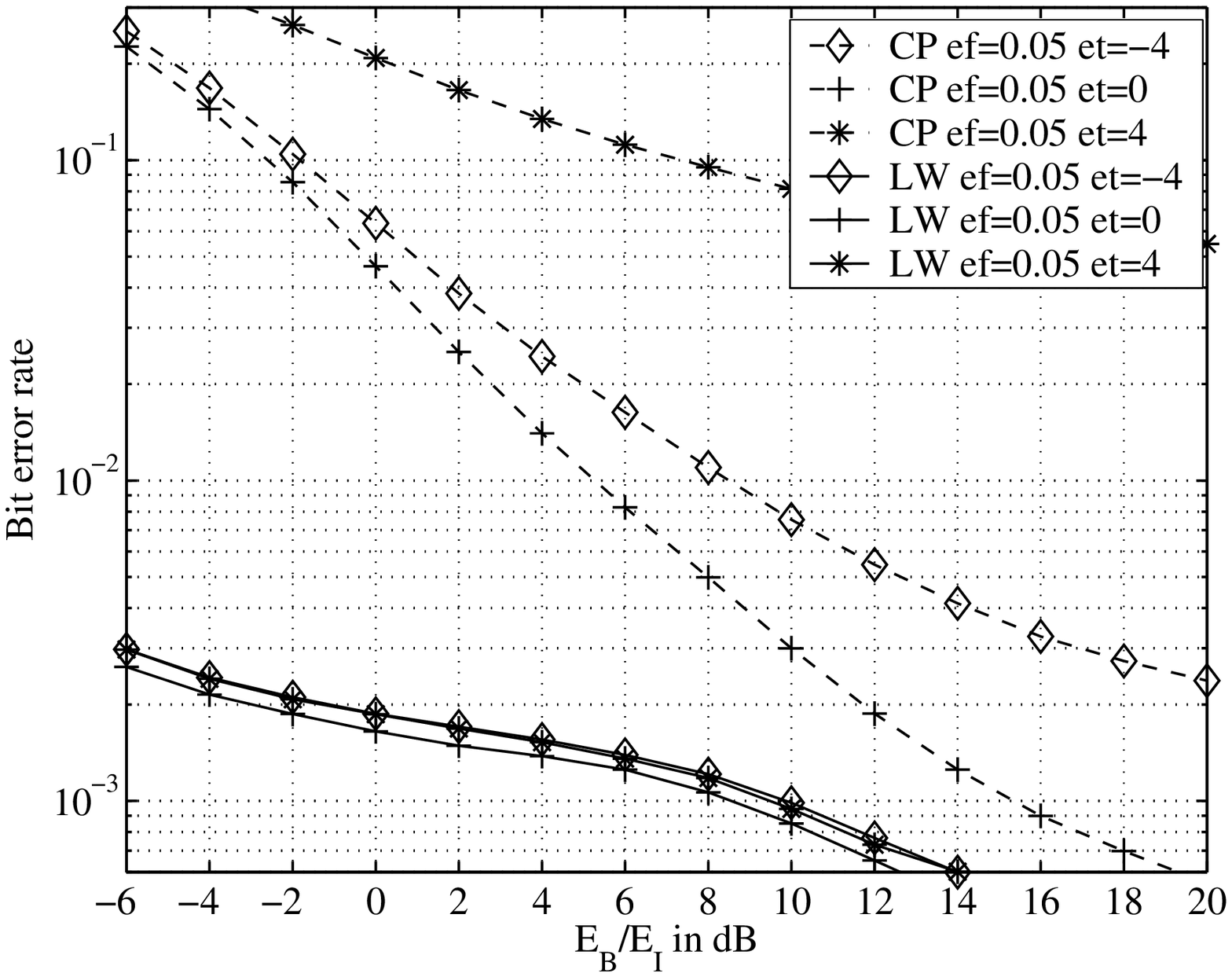} \\
a) & & b)
\end{tabular}}
\caption{\sl Measured bit error rates as a function of 
signal-to-interference ratios, $E_B/E_I$, for $N=36$-channel OFDM 
with $K=40$-sample frame interval in transmissions over a $3$-tap
Rayleigh channel
in the presence of narrowband interference, timing mismatch ($et$), and
frequency offset ($ef$). 
Two schemes are used: the cyclic prefix scheme  (CP)
and OFDM based on the waveform shown in Figure \ref{fig:cable}a) (LW). 
Frequency offset
($ef$) and timing mismatch ($et$) are set to the values given in the
legend. 
a) $ef=0$ - no frequency
offset.
b) $ef=0.05$ - frequency offset is $5\%$ of the bandwidth of one tone.}
\label{fig:ing40}
\end{figure}

The waveform shown in Figure~\ref{fig:cable}a) was used in simulations
of OFDM transmission 
in a narrowband interference scenario. Bit error rates were  measured as
a function of signal-to-interference ratio and are shown in
Figure~\ref{fig:ing40}.
We considered $N=36$-channel OFDM with  frame interval of
$K=40$ samples. Accordingly, the cyclic prefix was set to $K-N=4$
samples. The simulated channel
had $L=3$ Rayleigh distributed taps. 
The average powers of the taps in the ascending order were
$0$, $-3$, and $-6$ dB. Encoding and channel
estimation were performed in the same way as in the simulations
reported in the above. We introduced also additive white Gaussian noise,
the level of which was kept at $11$dB SNR, and in addition
to that we
placed a narrowband interferer with the bandwidth of one
subcarrier in the middle between the $17^{th}$ and $18^{th}$ OFDM
tone. The measured bit error rates of OFDM with the long waveform
(LW) and the cyclic prefix scheme (CP) as a function of
the $E_B/E_I$ values, where $E_I$ is the mean power of the
interferer, are shown in Figure~\ref{fig:ing40}
for different  frequency offset and
timing mismatch values. The results demonstrate 
that  the long waveform achieves 
a significantly lower bit error
rates in the presence of a narrowband interferer than the cyclic prefix 
scheme. Furthermore,
the larger the timing mismatch, frequency offset or the power of the
narrowband interferer, the more the performance of the cyclic prefix scheme
suffers. The scheme based on the long waveform, on the other hand, shows only
a minor increase in the bit error rate when the impairments and interference 
aggravate.

\section{Conclusion}

This paper presented a complete parameterization of OFDM modulating waveforms, {\sl i.e. }
orthonormal Weyl-Heisenberg sets in $\ell^2(\Z)$, and a complete
parameterization of corresponding biorthogonal demodulating waveforms. Several design 
examples are provided
and applied in typical transmission scenarios.
Simulations demonstrated a significant potential of long waveforms parameterized here
 for improving the robustness of OFDM to frequency offset, 
timing mismatch and narrowband interference.

\section*{Acknowledgement}
The authors are grateful to B.\ Vu\v{c}eti\'c for her support during
the course of this project.
The first author would also like to thank  L.\ J.\ Cimini for teaching him
principles of OFDM.


\begin{thebibliography}{00}

\bibitem{chang}
R.\ W.\ Chang. Synthesis of Band-Limited Orthogonal Signals for
Multichannel Data Transmission. {\sl Bell Syst. Tech. J.}
Vol.\ 45, pp.\ 1775-1796, Dec.\ 1966.

\bibitem{len}
L.\ J.\ Cimini. Analysis and Simulation of a Digital Mobile Channel
Using Orthogonal Frequency Division Multiplexing. {\sl IEEE
Trans.\ Commun.} Vol.\ 33, No.\ 7, pp.\ 665--675,
July 1985.

\bibitem{keller}
T.\ Keller and L.\ Hanzo, ``Adaptive Multicarrier Modulation:
A Convenient Framework for Time-Frequency Processing in
Wireless Communications,'' {\sl Proc.\ IEEE},
Vol.\ 88, No.\ 5, May 2000, pp.\ 611--640.

\bibitem{nee}
R.\ van Nee and R.\ Prasad.
\newblock {\em OFDM for Wireless Multimedia Communications.}
\newblock Artech House Publishers,  2000.

\bibitem{cherubini}
G. Cherubini, E. Eleftheriou, S. \"Olcer, and J. M. Cioffi.
\newblock Filter Bank Modulation Techniques for Very High Speed
Digital Subscriber Line.
\newblock {\em IEEE Commun.\ Magazine,}
\newblock Vol.\ 38, pp.\ 98--104, May 2000.

\bibitem{peled}
A.\ Peled and A.\ Ruiz. Frequency Domain Data Transmission Using
Reduced Computational Complexity Algorithms, in {\sl Proc. ICASSP'80}, 
Vol.\ 5, pp.\ 964--967, Apr.\ 1980.


\bibitem{poll}
T.\ Pollet, M.\ van Bladel, and M.\ Moeneclaey.
\newblock BER Sensitivity of OFDM Systems to Carrier Frequency
Offset and Wiener Phase Noise.
\newblock {\em IEEE Trans.\ Commun.}
\newblock Vol.\ 43, No.\ 2/3/4, pp.\ 191--193, Feb.--Apr. 1995.

\bibitem{ppv}
P.\ P.\ Vaidayanathan.{\sl Multirate Systems and Filter Banks}.
Prentice Hall, Englewood Cliffs, New Jersey, 1993.


\bibitem{tzannes}
S.\ D.\ Sandberg and M.\ A.\ Tzannes, ``Overlapped Discrete Multitone
Modulation for High Speed Copper Wire Communications,'' {\sl
IEEE J. Selected Areas in Communications,} Vol.\ 13, No.\ 9,
pp.\ 1571-1585, Dec.\ 1995.

\bibitem{holte96}
A. Vahlin and N. Holte, ``Optimal finite duration pulses for 
  OFDM,'' {\sl IEEE Trans.\  Commun.} Vol.\ 44, No.\ 1, pp.\ 10--14, 
Jan.\ 1996. 

\bibitem{holte97}
P.\ K.\ Remvik and N.\ Holte, ``Carrier frequency offset
robustness for OFDM systems with different pulse shaping filters,''   
{\sl Proc.\ IEEE GLOBECOM-97}, Vol.\ 1, pp.\ 11-15, Nov.\ 1997. 

\bibitem{hass}
R.\ Haas and J.\ C.\ Belfiore, ``A time-frequency well-localized 
pulse for multiple carrier transmission,'' {\sl Wireless Personal 
Commun.} Vol.\ 5, No.\ 1, pp.\ 1-18, July 1997. 

\bibitem{hleiss}
R.\ Hleiss, P.\ Duhamel, and  M.\ Charbit, ``Oversampled OFDM Systems,''
in {\sl Proc.\ DSP'97}, Vol.\ 1, pp.\ 329--332, July 1997.


\bibitem{kozek}
W.\ Kozek and A.\ F.\ Molisch, ``Nonorthogonal pulseshapes for 
multicarrier communications in doubly dispersive channels,'' 
{\sl IEEE J.\ Sel.\ Areas Commun.}, Vol. 16, No.\ 8, pp.\ 1579-1589, Oct.\ 1998.

\bibitem{scaglione1}
A.\ Scaglione, G.\ B.\ Giannakis and S.\ Barbarossa, 
"Redundant filterbank precoders and equalizers part I: unification and optimal design," 
{\sl IEEE Trans.\ on Signal Processing,} Vol.\ 47, No.\ 7, pp.\ 1988-2006, July 1999.

\bibitem{scaglione2}
A.\ Scaglione, G.\ B.\ Giannakis and S.\ Barbarossa, 
"Redundant filterbank precoders and equalizers part II: blind channel estimation, synchronization
and direct equalization," 
{\sl IEEE Trans.\ on Signal Processing,} Vol.\ 47, No.\ 7, pp.\ 2007-2022, July 1999.

\bibitem{cvetkovic99}
Z.\ Cvetkovi\'c, ``Modulating Waveforms for OFDM'' in
{\sl Proc.\ ICASSP'99}, Vol.\ 5, pp.\ 2463--2466, Mar.\ 1999.

\bibitem{bolcskei99}
H.\ B\"{o}lcskei, ``Efficient design of pulse shaping filters for 
OFDM systems,'' Proc.\ SPIE Wavelet Applications in Signal
and Image Processing VII 1999, pp.\ 625-636. 


\bibitem{cvetkovic00}
Z.\ Cvetkovi\'c, ``OFDM with Biorthogonal Demultiplexing,'' in
{\sl Proc.\ ICASSP'00}, Vol.\ 5, pp.\ 2517--2520, June 2000. 

\bibitem{lin}
Y.-P.\ Lin and S.-M.\ Phoong, ``ISI-Free Filterbank Transceivers for
Frequency-Selective Channels,'' {\sl IEEE Trans.\ Signal Processing},
Vol.\ 49, No.\ 11, pp.\ 2648--2658, Nov.\ 2001.

\bibitem{cherubini02}
G.\ Cherubini, E.\ Eleftheriou, S.\ \"{Olcer},
``Filtered Multitone Modulation fir Very High-Speed Digital
Subscriber Lines,'' {\sl IEEE J.\ Sel.\. Areas Commun.}, Vol.\ 20,
No.\ 5, pp.\ 1016--1028, June 2002. 

\bibitem{siohan}
P.\ Siohan, C.\ Siclet  and N.\ Lacaille, ``Analysis and Design
of  OFDM/OQAM systems based on the filter banks theory'',  {\sl IEEE 
Trans.\  Signal Processing,} Vol. 50, No.\ 5, pp.\ 1170-1183, May 2002.

\bibitem{benvenuto}
N. Benvenuto, S. Tomasin, and L. Tomba.
\newblock Equalization Methods in OFDM
and FMT Systems for Broadband Wireless Communications.
\newblock {\em IEEE Trans.\ Commun.,}
\newblock Vol.\ 50, No.\ 9, pp.\ 1413--1418, Sep.\ 2002.

\bibitem{strohmer}
T. Strohmer and S. Beaver, ``Optimal OFDM design for 
 time-frequency dispersive channels,'' {\sl IEEE Trans.\  Commun.}, 
Vol.\ 51, No.\ 7, pp.\ 1111-1122, July 2003. 


\bibitem{bolcskei03}
H.\ B\"{o}lcskei, P. Duhamel, and R. Hleiss, ``Orthogonalization of
OFDM/OQAM pulse shaping filters using the discrete Zak 
transform,'' Signal Processing, vol. 83, pp.\ 1379-1391, July 
2003.


\bibitem{phoong}
S.-M.\ Phoong, Y.\ Chang, and C.-Y.\ Chen, ``DFT Modulated Filterbank 
Transceivers for Multipath Fading Channels,'' {\sl IEEE Trans.\ Signal 
Processing}, Vol.\ 53, No.\ 1, pp.\ 182--192, Jan.\ 2005.

\bibitem{siclet} 
C.\ Siclet, P.\ Siohan, and D.\ Pinchon, ``Perfect Reconstruction 
Conditions and Design of Oversampled DFT-Modulated Transmultiplexers,'' 
{\sl EURASIP J. Applied Signal Processing,} Vol.\ 2006, Article ID 15756, 
pp.\ 1--14, 2006

\bibitem{matz}
G.\ Matz, D.\ Schafhuber, K.\ Gr\"{o}chenig, M.\ Hartmann, F.\ Hlawatsch, 
"Analysis, Optimization, and Implementation of Low-Interference Wireless 
Multicarrier Systems,'' {\sl IEEE Trans.\  Wireless Commun.}, 
Vol.\ 6, No.\ 5, pp.\ 1921--1931, May 2007. 

\bibitem{jung}
P.\ Jung and G.\ Wunder, "The WSSUS Pulse Design Problem in 
Multicarrier Transmission,'' {\sl IEEE Trans.\ Commun.},
Vol.\ 55, No.\ 10, pp.\ 1918--1928, Oct.\ 2007.


\bibitem{cvetkovic98}
Z.\ Cvetkovi\'c and M.\ Vetterli, ``Tight Weyl-Heisenberg Frames in 
$\ell^2(\Z)$,'' {\sl IEEE Trans.\ Signal Processing}, Vol.\ 46, No.\ 5,
pp.\ 1256--1259, May 1998.


\bibitem{shannon}
C.\ Shannon, ``Communication in the Presence of Noise,'' {\sl Proc.\ IRE,}
Vol.\ 37, pp.\ 10--21, 1949.

\bibitem{martin}
M.\ Vetterli and J.\ Kova\v{c}evi\'{c}. {\sl Wavelets and Subband
Coding.} Prentice-Hall,
Englewood Cliffs, New Jersey, 1995.

\bibitem{janssen}
A.\ J.\ E.\ M.\ Janssen, ``Duality and Biorthogonality for
Weyl-Heisenberg Frames,'' {\sl J.\ Fourier Analysis and Applications},
Vol.\ 1, No.\ 4, pp.\  403--436, Nov.\ 1995.

\bibitem{daubechies}
I.\ Daubechies, H.\ J.\ Landau, and Z.\ Landau, ``Gabor Time-Frequency Lattices
and the Wexler-Raz Identity,'' {\sl J.\ Fourier Analysis and 
Applications}, Vol.\ 1, No.\ 4, pp.\ 437--478, Nov.\ 1995.

\bibitem{cvetkovic00_1}
Z.\ Cvetkovi\'c, "On Discrete Short-Time Fourier Analysis,"
 {\sl IEEE Trans.\ Signal Processing}, Vol.\ 48, No.\ 9,
pp.\ 2628--2640, Sept.\ 2000.


\bibitem{vcmt_prop}
D.\ Andelman,  ``Variable Constellation
Multitone (VCMT) Proposal,'' Ultracom Communications Inc.,
IEEE 802.14a/98-013 document, June 1998.

\bibitem{blahut}
R.\ E.\ Blahut, {\sl Fast Algorithms for Digital Signal
Processing}. Addison-Wesley Publishing Company, Inc.\ 1985.

\bibitem{alan}
H.\ Park. {\sl A Computational Theory of Laurent Polynomial Rings and
Multidimensional FIR Systems,} PhD Thesis, Department of Mathematics,
UC Berkeley, May 1995.

\bibitem{alan2}
H.\ Park. Parahermitian Modules and Paraunitary Groups. {\sl preprint}, 1998.

\bibitem{han}
F.-M.\ Han and X.-D.\ Zhang, "Wireless Multicarrier Transmission via
Weyl-Heisenberg Frames over Time-Frequency Dispersive Channels,"
{\sl IEEE Trans.\ Communications}, Vol.\ 57, No.\ 6, pp.\ 1721--1733, June 2009.

\end{thebibliography}





\end{document}